\documentclass[aps,prb,reprint,amsmath]{revtex4-2}
\usepackage{color}
\usepackage[caption=false]{subfig}
\usepackage{graphicx}

\DeclareMathOperator{\tr}{Tr}
\newcommand{\eH}{\mathcal{H}}
\newcommand{\up}{\uparrow}
\newcommand{\down}{\downarrow}
\newcommand{\ep}{\epsilon}
\newcommand{\si}{\sigma}

\newcommand{\mG}{\mathbf{G}}
\newcommand{\tmG}{\tilde{\mG}}

\begin{document}

\title{The absence of Ginzburg-Landau mechanism for vestigial order in the normal phase above a two-component superconductor}

\author{P. T. How}
\email{pthow@outlook.com}
\affiliation{Institute of Physics, Academia Sinica, Taipei 115, Taiwan}
\author{S. K. Yip}
\email{yip@phys.sinica.edu.tw}
\affiliation{Institute of Physics, Academia Sinica, Taipei 115, Taiwan}
\affiliation{Institute of Atomic and Molecular Sciences, Academia Sinica, Taipei 115, Taiwan}

\date{\today}

\begin{abstract}

A two-component superconductor may hypothetically support a vestigial order phase above its superconducting transition temperature, with rotational or time-reversal symmetry spontaneously broken while remain non-superconducting.  This has been suggested as an explanation to the observed normal state nematicity of the nematic superconductor $M_x$Bi$_2$Se$_3$.  We examine the condition for this vestigial order to occur within Ginzburg-Landau theory with order parameter fluctuations, on both the nematic and chiral sides of the theory.  Contrary to prior theoretical results, we rule out a large portion of parameter space for possible vestigial order.  We argue that very extreme anisotropy is one prerequisite for the formation of a stable vestigial phase via this mechanism, which is likely not met in real materials.

\end{abstract}

\maketitle

\section{Introduction}

Superconductivity in doped topological insulator Bi$_2$Se$_3$ has captured much recent attention.  While the crystal is supposed to have D$_{3d}$ symmetry that sees \emph{three-fold} rotational symmetry in the basal plane, \emph{two-fold} anisotropy in the superconducting phase in the basal plane have been observed in experiments\cite{Matano2016-vr, Pan2016-ni, Asaba2017-vn, Shen2017-bq, Yonezawa2017-fr, Smylie2018-ab}.  See Yonezawa \cite{Yonezawa2018-to} for a review.  \emph{Nematic superconductivity} is a proposed explanation \cite{Fu2014-gm}.   More precisely, it has been proposed
that the superconducting order parameter belongs to a two-dimensional representation, and the energetics is such that, below the superconducting transition, the order parameter picks a state with spontaneously broken rotational symmetry (other than the other possibility where time reversal symmetry is broken, \emph{c.f.} the case for UPt$_3$ \cite{Sauls1994-ol, Joynt2002-xq}).

If the order parameter belongs to a two-dimensional representation, one expects an internal degree of freedom (rotation of the order parameter in this case) to reveal itself under suitable circumstances.  However, so far no experiments have convincingly shown this degree of freedom.  One may expect external stress can re-orient the order parameter \cite{How2019-re}, but an experiment at Argonne %
\footnote{
Talk by Kristin Willa at Spin Phenomena Interdisciplinary Center, ``Evidence for nematic superconductivity in superconducting doped topological insulator  Nb$_x$Bi$_2$Se$_3$ and Sr$_x$Bi$_2$Se$_3$'', available on https://www.youtube.com/watch?v=gVuzkKCU1xg
}
 turns out to be negative.  In a related experiment on multidomain sample at Kyoto \cite{ Kostylev2020-zb}, only changes of the relative sizes of the domains were found.  One might also expect that there should be special features in the upper critical field such as kinks as a function of the magnitude of the field \cite{Willa2018-uu} (\emph{c.f.} \cite{Hess1989-wi}) or angle in the plane \cite{Venderbos2016-fa}.  Neither has been reported so far and a recent experiment \cite{Bannikov2021-uh} specifically looking for these features was not able to find one.  Others \cite{Zyuzin2017-kn} and us \cite{How2020-qj} have predicted the existence of half quantum vortices or skyrmions (which are unique to multicomponent order parameters but absent in single component systems).  We have also investigated the special features in shear stress tensor due to the multi-dimensional nature of order parameter \cite{How2021-me}.   Experiments examining these predictions have not yet been reported.

A nematic superconducting order breaks both gauge and rotational symmetry.  In principle the symmetries are not necessarily broken at the same temperature.  A few years ago,  \cite{Hecker2018-bb} predicted that ``vestigial nematic order" can exist in this system:  as the temperature is lowered, the symmetry preserving normal state first makes a transition into a state with broken rotational symmetry, and only later gauge symmetry is broken, forming the nematic superconducting state.  This possibility is unique to a multi-component order parameter: a superconductor with an order parameter belonging to a one-dimensional representation, even if it is not s-wave, cannot exhibit this vestigial order.  Observation of this ``vestigial nematic state" would be a ``smoking gun" of this nature of the order parameter.  Electronic nematicity in the normal phase was first reported in \cite{Sun2019-kt}, and then \cite{Cho2020-ne} claimed to observe the double transitions, with a vestigial order phase sandwiched in between normal and superconducting phases. In particular, length change of the sample as a function of temperature or field was monitored.  A rapid and directional dependent change as a function of temperature above the superconducting transition was observed and interpreted by these authors as a step indicating a first order transition into a vestigial nematic-ordered state.  It is remarkable that the relative change in length is only of order $10^{-7}$, even smaller than the distortion from perfect D$_{3d}$  found at room temperatures from another group \cite{Kuntsevich2018-kk}.  We are therefore not sufficiently convinced by \cite{Cho2020-ne}: their result may simply reflect a  broadened superconducting transition due to, e.g., sample inhomogeneity.  We are thus motivated to consider the criterion for
vestigial order in more detail in the present paper.  Time-reversal-symmetry-breaking vestigial order associated with a superconductor was discussed in \cite{Bojesen2014-vj, Fischer2016-gt}.  Vestigial orders have been recently discussed in many other systems \cite{Fernandes2019-yv, Grinenko2021-cw}.

As mentioned above, a two-component order parameter can instead breaks the time-reversal symmetry and form a \emph{chiral} superconductor.  A unifying Ginzburg-Landau (GL) theory in two complementary regions of parameter space describes respectively chiral and nematic superconductors.  To be more specific, the sign of $\beta_2$ introduced in \eqref{Hi} controls the choice: negative for nematic and positive for chiral.  While only the nematic side of the theory is experimentally relevant to $M_x$Bi$_2$Se$_3$, we study both sides of the theory.  The relevance of chiral superconductivity in $M_x$Bi$_2$Se$_3$ is discussed in \cite{Yuan2017-la, Chirolli2018-zj,  Uematsu2019-ij}.

In this paper, we set out to find vestigial order for both the chiral and the nematic sides of the GL theory using a modified Luttinger-Ward (LW) approach, and find none.  For the region $\vert\beta_2/\beta_1\vert < 1/2$ ($\beta_1$ also given in \eqref{Hi}), we expect a second order transition from the normal phase to the appropriate (chiral or nematic) superconducting phase, \emph{without a vestigial order in between.}  For  $\beta_2/\beta_1 < -1/2$ (``deeply nematic'') or $\beta_2/\beta_1 > 1/2$ (``deeply chiral''), we predict a joint first order transition directly into the appropriate superconducting phase and \emph{no vestigial order} for physically relevant choices of parameters.  Our negative result contradicts \cite{Hecker2018-bb, Cho2020-ne}, but is not inconsistent with \cite{Fischer2016-gt}.  Comparison with these prior works will be made as we present our result.  Previously we made available a preprint \cite{How2022-qb} that erroneously predicted a vestigial nematic phase in the deep nematic region.  We will comment on the difference in section IV A.

The organization of this paper is as followed.  In Section II we introduce the effective Hamiltonian for the two-component order parameter, and also the concept of a vestigial order.  In Section III we give the formulas of the LW formalism with UV-divergence subtraction.  We report and discuss our results for the chiral and nematic cases in Sections IV and V, respectively.  Section VI is the conclusion.

\section{Theoretical Model}

The superconductors in question can all be described by a two-component complex order parameter field
\begin{equation}
\eta(\vec{r}) = \begin{pmatrix}
\eta_x(\vec{r}) \\ \eta_y(\vec{r})
\end{pmatrix}.
\end{equation}
As the notation suggests, under rotations about the $z$-axis, $\eta$ transforms just like a vector in the $xy$-plane.  Focusing on the case with trigonal symmetry, one can write down a phenomenological effective Hamiltonian density $\eH = \eH_K + \eH_i$, split into the kinetic part $\eH_K$ and the interaction $\eH_i$:
\begin{widetext}
\begin{align}
\beta \eH_K &=
	\alpha (\eta_j^* \eta_j)
	+ K_1 (\partial_i \eta_j)^* (\partial_i \eta_j)
	+ K_2  (\partial_i \eta_i)^* (\partial_j \eta_j)
	+ K_3  (\partial_i \eta_j)^* (\partial_j \eta_i)
	+ K_{zz} (\partial_z \eta_j)^* (\partial_z \eta_j)  \nonumber \\
	& \qquad + \frac{K'}{2} \left[ (\partial_z \eta_y^*)
		(\partial_x \eta_x - \partial_y \eta_y)
	+ (\partial_z \eta_x^*)  (\partial_x \eta_y + \partial_y \eta_x) + c.c. \right];
	\label{Hk} \\
\beta \eH_i &=
	\frac{\beta_1}{2} (\eta_i^* \eta_i) (\eta_j^* \eta_j)
	+ \frac{\beta_2}{2} (\eta_i \eta_i)^* (\eta_j \eta_j).
	\label{Hi}
\end{align}
\end{widetext}
Repeated indices $i$ or $j = x, y$ are summed over, and ``c.c.'' stands for complex conjugate.  The parameter $\alpha$ labels the temperature, as is usual in the GL theory.  We may also write $\alpha = \alpha' (T-T_0)$, where $T_0$ is the meanfield critical temperature of superconductivity, and we are interested only in the small-$\alpha$ limit.  $\beta = (k_B T)^{-1} \approx (k_B T_0)^{-1}$ is the usual inverted temperature, and can be regarded as approximately a constant.  Effective Hamiltonian density of this form has been adopted in the literature \cite{Venderbos2016-fa,Hecker2018-bb}, and the notation here is in line with our previous papers \cite{How2019-re,How2021-me}.  The gradient terms proportional to $K_{1,2,3}$ exhaust all allowed possibility in a completely cylindrically symmetric or a $D_6$ system, while $K'$ is additionally allowed by the lower $D_{3d}$ symmetry \cite{Barash1991-qg}.  The ``Fermi surface warping'' discussed in \cite{Akzyanov2020-zb} is one possible origin of such $K'$ term
\footnote{If one attempts to construct this effective theory starting from the approximate $k \cdot p$ electron Hamiltonian of $M_x$Bi$_2$Se$_3$ \cite{Zyuzin2017-kn}, then $K'$ (accidentally) vanishes.  This is an artifact of the $k \cdot p$ approximation.}.  
Thermodynamics is formally governed by the partition function $Z = \int \mathcal{D}\eta \mathcal{D}\eta^{\dag} e^{-\int \! d^3 r \, \beta\eH}$.

Stability requires $\beta_1 > 0$ and $\beta_2 > -\beta_1$, so that $\eH_i$ is bounded from below.  If $\beta_2 < 0$, the uniform meanfield ground state of $\eH$ has a finite and real $\eta$ up to an overall phase factor.  This is the nematic superconducting state that breaks both global $U(1)$ and rotational symmetry.  On the other hand, if $\beta_2 > 0$, the meanfield ground state favors $\eta_x = \pm i \eta_y$.  This is the chiral superconducting state, which is invariant under spatial rotation, but breaks the time-reversal symmetry along with the global $U(1)$ symmetry.

It turns out that the physics is more intuitively represented in the alternative basis
\begin{equation}
\Phi =
	\begin{pmatrix}
		\phi_{\up} \\
		\phi_{\down}		
	\end{pmatrix}
	\equiv	
	\frac{1}{\sqrt{2}}
	\begin{pmatrix}
		\eta_x + i\eta_y \\
		\eta_x - i\eta_y
	\end{pmatrix}.
\end{equation}
As the notation suggests, we will adopt the analogy of $\Phi$ being a (pseudo-)spin-$1/2$ object.  The original $\eta$ is related to $\Phi$ by a change of spin quantization axis in this language.  We will take the thermodynamic limit, pass to the Fourier space, and rescale the momenta so that $\eH_K$ appears in a much more pleasing form:
\begin{equation}
\begin{split}
\beta \eH_K &= \int_p \Phi^{\dag}(p)
			\left( \ep_0(p)\, \si_0 + \ep_x(p)\, \si_x + \ep_y(p)\, \si_y \right)
				\Phi(p); \\
\ep_0(p) &= p_x^2 + p_y^2 + p_z^2 + \alpha;\\
\ep_x(p) &= C_1 \left(p_x^2 - p_y^2 \right)\!\!/2 + C_2 \, p_z p_y; \\
\ep_y(p) &= C_1 \, p_x p_y + C_2 \, p_z p_x;\\
C_1 =& \frac{K_2 + K_3}{2K_1+K_2+K_3}; \quad C_2 = \frac{K'}{\sqrt{K_z(2K_1+K_2+K_3)}}.
\end{split}
\label{HkPhi}
\end{equation}
We introduce the shorthand notation $\int_p \equiv \int d^3p/(2\pi)^3$; $\si_{x,y,z}$ are the usual Pauli matrices, and $\si_0$ is the identity matrix.  Terms without derivatives are not affected by the rescaling.  In the pseudospin language, $\ep_x$ and $\ep_y$ are effectively spin-orbit coupling terms.

In terms of $\Phi$, $\eH_i$ can be written as
\begin{equation}
\begin{split}
\beta \eH_i
&=
	\frac{g_1}{2} \left(\vert\phi_{\up}\vert^4 + \vert\phi_{\down}\vert^4 \right)
	+ g_2 \, \vert\phi_{\up}\vert^2 \vert\phi_{\down}\vert^2 \\
&=
	\sum_{i = x,y,z} \frac{\lambda_i}{2} \left(\Phi^{\dag}\si_i \Phi\right)^2
\end{split}
\label{HiPhi}
\end{equation}
where $g_1 = \beta_1$ and $g_2 = \beta_1 + 2\beta_2$.  The stability requirement is $g_1 > 0$, $g_2 > - g_1$.  The meanfield nematic (chiral) superconducting phase requires $g_1$ greater (smaller) than $g_2$.  In the second line, the coefficients are $\lambda_x = \lambda_y = (g_1+g_2)/2$ and $\lambda_z = g_1$.  This form is often more useful in calculation.

In the meanfield description, the global $U(1)$ symmetry and rotational (time-reversal) symmetry are both spontaneously broken at the critical temperature for a nematic (chiral) superconductor.  When fluctuation is included, however, \emph{a priori} there is no reason for everything to occur at once.  Indeed, a non-zero expectation value of $\langle\Phi^{\dag} \sigma_z \Phi\rangle$ breaks the time-reversal symmetry, but still respects the global $U(1)$ symmetry.  For the rotational symmetry, $\langle\Phi^{\dag} \sigma_{x,y} \Phi\rangle$ is the analogous quantity.  Either may be non-vanishing while $\langle \Phi\rangle$ itself remains zero.  Ref \cite{Hecker2018-bb} proposed the existence of $\langle\Phi^{\dag} \sigma_{x,y} \Phi\rangle \neq 0$ ordering mediated by superconducting fluctuation above $T_c$ in $M_x$Bi$_2$Se$_3$, giving rise to the so-called vestigial nematic order.  Historically, the idea of such bilinears can be found in the study of the ``metallic superfluid'' \cite{Babaev2004-op} and ``super-counter-fluid'' \cite{Kuklov2004-ib}, and subsequently seen in several other models for multi-component superfluid or superconductor \cite{Kuklov2008-oz, Herland2010-di, Bojesen2013-lt, Bojesen2014-vj}%
\footnote{
Subsequent discussions are often framed in terms of the ordering in \emph{relative phase} between field components, and thus bear less superficial resemblance to the present scenario.  And of course the models and the mechanisms discussed there are very different.
}.

In the spin-$1/2$ metaphor, the vestigial nematic order discussed here is a spin order in the $xy$-plane: a superposition of up and down spin.  The fact that $\langle \Phi^{\dag}\si_{x,y}\Phi\rangle \neq 0$ requires $\langle \phi^{*}_{\up} \phi_{\down}\rangle \neq 0$.  Equation \eqref{HiPhi} suggests that the order is favored only when $g_2 < 0$, and one expect no vestigial order at all for $g_1 > g_2 > 0$.  This criterion contradicts the previous theoretical result \cite{Hecker2018-bb, Cho2020-ne}, and is indeed the initial motivation of the present work.  Though it will presently be shown that $g_2 < 0$ alone is insufficient to stabilize the vestigial nematic phase.

\begin{figure}
\includegraphics[scale=1]{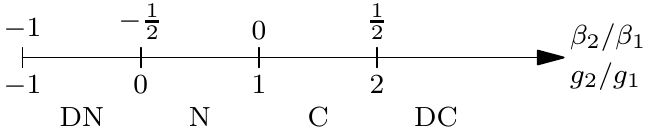}
\caption{\label{parameter} A visual guide for translating between $\beta_2/ \beta_1$ and $g_2/g_1$ (note that the relation is non-linear).  The ratios control critical behavior of the system.  The labels DN, N, C, DC are deep nematic, nematic, chiral, deep chiral, respectively.}
\end{figure}

On the chiral side, we will later deduce a similar necessary (but insufficient) condition $g_2 > 2g_1$ for the vestigial phase, but we do not see a simple way to read this off from \eqref{HiPhi}.  These $g_2 < 0$ and $g_2 > 2g_1$ regions corresponds to the deep nematic and deep chiral regions introduced in the previous section; see Fig \ref{parameter}.

We note that the deep nematic region $g_2 < 0$, while not obviously forbidden, seems not easily attained, either.  Calculations of these GL coefficients from various microscopic models are done in supplementary materials of \cite{Venderbos2016-cv, Zyuzin2017-kn}.  It appears to us that, of all the models discussed in these two references, none exhibits a negative $g_2$%
\footnote{
Ref \cite{Venderbos2016-cv} 
wrote the quartic terms of the free energy in a different form: their $B_1$ and $B_2$ are proportional to our $\beta_1 + \beta_2$ and $-\beta_2$, respectively.  The sign of our $g_2$ is thus the same as that of $B_1 - B_2$ in their supplementary material.  It seems to us that this quantity is always non-negative for every model discussed there.
}.

Unfortunately, a treatment using $\langle\Phi^{\dag} \sigma_i \Phi\rangle$ as the order parameter is far from straightforward.  \cite{Hecker2018-bb} attempted this but was forced to employed a Hubbard-Stratonovich decoupling which fails to account for all scattering channels, and this is likely the main cause of their error.  (See the appendix of \cite{Fischer2016-gt} and the present authors' preprint \cite{How2022-qb} for further discussion.)  Here we appeal to the spin metaphor again: a spin order and an effective Zeeman splitting along the direction of the order mutually imply each other.  We therefore look instead at the \emph{self energy} of $\Phi$, and employ the Zeeman part as the order parameter for the vestigial phase.

\section{Renormalized Bosonic Self energy Method}

The celebrated LW functional \cite{Luttinger1960-el} is the natural method to treat self energy in a self-consistent way.  To make it work for the bosonic $\Phi$ field, we have to make a simple extension so that the UV divergence of a bosonic field theory is handled properly.  Well-known extra complication for this divergence-removal process is present in $d=4$ \cite{Baym1977-lm, Van_Hees2001-hl, Blaizot2004-ae, Berges2005-yz}, but the problem at hand effectively lives in $d=3$, and the process is very straightforward.  We will state the formulas here, and interested reader are referred to the appendix for a more detailed discussion.

In $d=3$, the temperature parameter $\alpha$ is the only quantity that receives UV-divergent correction.  Following the standard procedure of perturbative renormalization \cite{Fradkin2021-zu}, a counter term $\delta\alpha \vert\Phi\vert^2$ is added to $\beta\eH$ to cancel the the divergence.  This counter term is determined order-by-order in perturbation theory in any preferred renormalization scheme, and without reference to any self-consistent condition yet.

We assume the theory is translationally invariant and take the thermodynamic limit.  Let $\mG(k)$ be the (variational) momentum-space matrix propagator of the interacting theory:
\begin{widetext}
\begin{equation}
(2\pi)^3 \delta^{(3)}(k-q)\, [\mG(k)]_{ij}
	\equiv
	\int\!\! d^3 \! x \!\! \int\!\! d^3\! y
	\; e^{i (k \cdot x - q \cdot y)} \, \langle\phi_i^{*}(x) \phi_j(y) 				\rangle
\end{equation}
The free energy density functional $\Omega[\mathbf{G}]$ now reads
\begin{equation}
\begin{split}
\beta \Omega[\mG] &=
	\int_k \tr \log \mG(k)^{-1} - \int_k \tr\! \left\lbrace\left[\mG(k)^{-1} - \mG_0(k)^{-1} \right] \mG(k)\right\rbrace \\
	& \qquad + \delta\alpha \int_k\! \tr \mG(k) \, + \, \Phi_{LW}[\mG].
\end{split}
\label{LWnoSSB}
\end{equation}
\end{widetext}
We have introduced the shorthand $\int_k = \int d^3 k/(2\pi)^3$, and $\tr$ here refers exclusively to the matrix trace.  $\mG_0$ is the free propagator.  The LW functional $\Phi_{LW}[\mG]$ is defined as usual to be the generating functional of two-particle irreducible self energy diagrams \cite{Luttinger1960-el, Hugel2016-bx}.  The first line of \eqref{LWnoSSB} is the non-interacting part, and the second line vanishes in the non-interacting limit.  This free energy itself still contains a UV-divergent additive constant, but the physically relevant quantity $(\Omega[\mG] - \Omega[\mG_0])$ is UV-finite to all order in perturbation theory.  This claim will be shown explicitly to one-loop order in the subsequent calculation, and the general argument is presented in the appendix.

Superfluid order can be incorporated into this formalism, too.  We restrict ourselves to a uniform order $\bar{\Phi}\equiv \langle \Phi \rangle \neq 0$ here, and let the fluctuating part be $\tilde{\Phi} \equiv \Phi - \bar{\Phi}$.  Now that $\bar{\Phi}$ breaks the $U(1)$ symmetry, we have to adopt the Nambu spinor notation to effectively account for four real scalar degrees of freedom:
\begin{equation}
\Psi =
\begin{pmatrix}
\psi_{11} \\ \psi_{12}\\ \psi_{21}\\ \psi_{22}
\end{pmatrix}
\equiv
\begin{pmatrix}
\tilde{\phi}_{\up} \\ \tilde{\phi}^{*}_{\up} \\ \tilde{\phi}_{\down} \\ \tilde{\phi}_{\down}^{*}
\end{pmatrix}.
\end{equation}
Expanding the original $\eH$ yields additional effective self energy and cubic interaction for $\Psi$; let us put
\begin{equation}
\beta\eH[\Phi] = \beta\eH[\bar{\Phi}] + \beta\eH[\tilde{\Phi}] +  \frac{1}{2} \Psi^{\dag} \mathbf{M} \Psi + \frac{1}{3!} N_{ijk} \psi_i \psi_j \psi_k
\end{equation}
where the coefficients $\mathbf{M}$ and $N_{ijk}$ are, of course, dependent on $\bar{\Phi}$ (and $\bar{\Phi}^{\dag}$).

Let $\tmG$ be the $4 \times 4$ matrix propagator of the $\Psi$ field, and $\tmG_0$ the non-interacting limit of that.  The free energy density with possible superconducting order is
\begin{widetext}
\begin{equation}
\begin{split}
\beta \Omega[\tmG] &=
	\frac{1}{2}\int_k \tr \log \tmG(k)^{-1} - \frac{1}{2}\int_k \tr\! \left\lbrace\left[\tmG(k)^{-1} - \tmG_0(k)^{-1} \right] \tmG(k)\right\rbrace \\
	& \qquad + \frac{1}{2}\int_k\! \tr\left[(\delta\alpha + \mathbf{M}) \tmG(k)\right] \,
		+ \, \tilde{\Phi}_{LW}[\tmG]
		+ \eH[\bar{\Phi}] + \delta\alpha\vert\bar{\Phi}\vert^2.
\end{split}
\label{LWwithSSB}
\end{equation}
\end{widetext}
The LW functional $\tilde{\Phi}_{LW}$ is defined as the generator of 2PI self energy diagrams for the $\Psi$ field, taking into account the effective cubic interaction $\frac{1}{3!} N_{ijk} \psi_i \psi_j \psi_k$.

Great care must be taken, however, when interpreting the result whit superfluid order.  It has been known \cite{Baym1977-lm, Hugel2016-bx} that similar approaches (within Hartree-Fock-like approximation) always predict first order transitions even when second order ones are expected on symmetry ground, and the solutions always weakly violate the Goldstone theorem in the superfluid phase; it is believed that these methods do not adequately handle the strong critical fluctuation when the spectrum is nearly gapless.  Unfortunately our $d=3$ renormalized variant turns out to be no different.  Such artifacts will be discussed in more detail when they are encountered.

\section{The Chiral side}

\subsection{Decoupled limit}
\label{4a}
Let us first focus on the story of the chiral side of the theory ($g_1 < g_2$) given the much cleaner algebra, even though $M_x$Bi$_2$Se$_3$ is on the nematic side.  Most of the physics is in fact very similar on either sides.  We will start with the decoupled limit with $C_1 = C_2 = 0$, so called because the effective spin-orbit coupling vanishes.  The gradient term now has full rotational invariance, but the interaction still distinguishes the $z$-direction from the rest, so the theory enjoys full cylindrical symmetry.  We will confine ourselves to $\alpha > 0$.  In the subsequent discussion, apart from the (self-explanatory) superconducting and vestigial phases, we will refer to the high temperature phase without any ordering as the ``symmetric phase''.

\begin{figure}
\subfloat[\label{vacuum2}]{\includegraphics[width=0.2\textwidth]{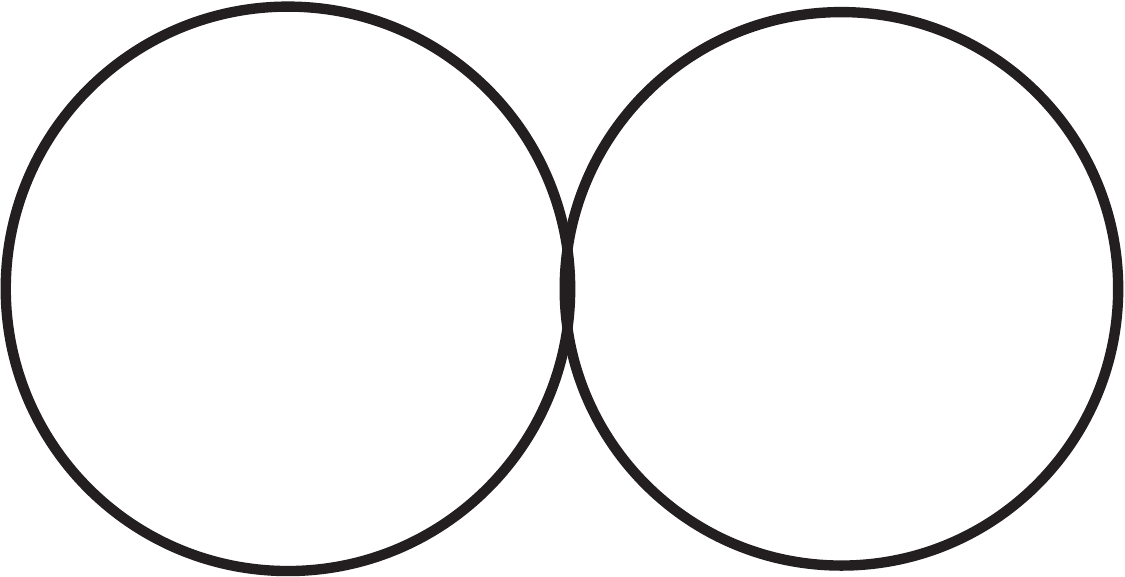}}\qquad \quad
\subfloat[\label{2PI1}]{\includegraphics[width=0.2\textwidth]{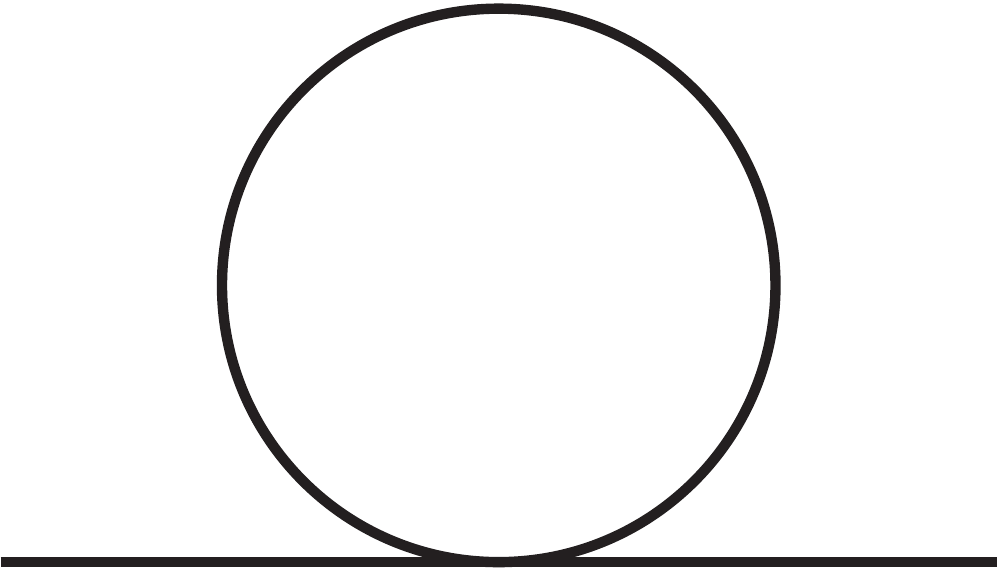}}
\caption{(a) The two-loop Hartree-Fock vacuum diagram contributing to the LW functional $\Phi_{LW}$.  (b) The associated one-loop proper self energy diagram.}
\end{figure}

The main approximation here is the truncation of $\Phi_{LW}$ to the Hartree-Fock term represented by the Feynman diagram Fig \ref{vacuum2}.  In terms of the full propagator $\mG$, it reads
\begin{equation}
\begin{split}
\Phi_{LW} \approx
	\sum_{i = x, y, z} & \frac{\lambda_i}{2} \Bigg\lbrace \left[\int_k \tr(\si_i \mG(k))\right]^2 \\
		&+ \tr\left[\si_i \left(\int_p \mG(p)\right) \si_i \left(\int_k \mG(k)\right)\right]
		\Bigg\rbrace.
		\end{split}
\label{HartreeFock}
\end{equation}
where the second form of \eqref{HiPhi} is used.  We have not specify our ansatz for $\mG$, but any sensible choice would make \eqref{HartreeFock} UV-divergent.  We will simply assume that the integrals are regularized in some suitable scheme and press on.

The non-interacting propagator is identified as $\mG_0(k) = (k^2 + \alpha)^{-1} \si_0$.  The proper self energy diagram generated by \eqref{HartreeFock} is depicted in Fig \ref{2PI1}.  It represents a momentum-independent energy shift.  We therefore adopt the ansatz
\begin{equation}
\mG(k)^{-1} = (k^2 + \alpha + h_0)\si_0 + \vec{h} \cdot \vec{\si},
\end{equation}
where $h_0$ and $\vec{h} = (h_x, h_y, h_z)$  are the variational parameters.  The traceless part $\vec{h}\cdot\vec{\si}$ is the induced Zeeman splitting: non-vanishing $h_x$ or $h_y$ indicates vestigial nematic order, while a non-zero $h_z$ implies vestigial chiral order.  It is necessary to allow for $h_0$ variation: while $\alpha$ is fixed by the physical temperature, it is the average energy gap of the \emph{symmetric phase only}.  Any other solution can in principle have a different value of average gap at the same temperature, reflected by a $h_0 \neq 0$.
\footnote{
The omission of the $h_0$ variation in our previous preprint \cite{How2022-qb} was the single mistake that led to its erroneous conclusion.  It amounted to an extra constraint placed on the system.   With $h_0$ artificially set to zero here, one can recover the result of \cite{How2022-qb}.
}

We opt for the following condition to specify the renormalization counter term $\delta\alpha$: in the symmetric phase, the one-loop correction Fig \ref{2PI1} should be exactly canceled by the counter term, leaving the renormalized value $\alpha$ unchanged.  This is accomplished by choosing
\begin{equation}
\begin{split}
\delta \alpha &= -(2 g_1 + g_2) \int_k \frac{1}{k^2 + \alpha} \\
&= -(\lambda_x + \lambda_y + \lambda_z) \int_k \frac{1}{k^2 + \alpha},
\end{split}
\end{equation}
also regularized with the same suitable scheme used in \eqref{HartreeFock}.

With all the necessary ingredients in place, the free energy density $\Omega$ given in \eqref{LWnoSSB} can be calculated up to the Hartree-Fock approximation.  As mentioned earlier, one needs to discard a UV-divergent additive constant, and then the expression is manifestly UV-finite.  One recognizes that the gap matrix $[(\alpha + h_0)\si_0 + \vec{h}\cdot\vec{\si}]$ has eigenvalues $\alpha + h_0 \pm \vert\vec{h}\vert$, and defines the quantities
\begin{equation}
\begin{split}
m_1 &= \sqrt{\alpha + h_0 + \vert\vec{h}\vert} \geq 0,\\
m_2 &= \sqrt{\alpha + h_0 - \vert\vec{h}\vert} \geq 0
\end{split}
\label{GapEigenvalues}
\end{equation}
to be the \emph{square root} of the eigenvalues.  Now the free energy density relative to the symmetric phase can be conveniently written as
\begin{widetext}
\begin{equation}
\begin{split}
\beta(\Omega - \Omega_{s})
&= 
	\frac{1}{12\pi} (
		m_1^3 + m_2^3 )
	- \frac{\alpha}{4\pi} (
		m_1
		+ m_2)
	+ \frac{(2g_1 + g_2)}{64\pi^2}
		(m_1 + m_2 - 2\sqrt{\alpha})^2 \\
&\qquad + \frac{1}{64 \pi^2 \vert \vec{h}\vert^2}
		\left[ g_2 (h_x^2 +h_y^2) + (2g_1-g_2) h_z^2\right]
		(m_1 - m_2)^2 + \frac{1}{3\pi} \alpha^{3/2},
\end{split}
\label{MasterFreeEnergy}
\end{equation}
\end{widetext}
where $\Omega_{s}$ is the free energy density of the symmetric solution $m_1 = m_2 = \sqrt{\alpha}$.

\begin{figure}
\subfloat[\label{landscapeHigh}]{\includegraphics[scale=1]{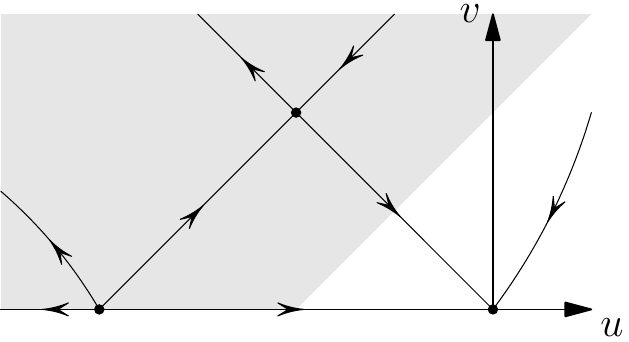}} \qquad
\subfloat[\label{landscapeMid}]{\includegraphics[scale=1]{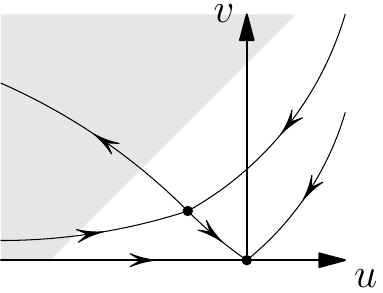}} \quad
\subfloat[\label{landscapeLow}]{\includegraphics[scale=1]{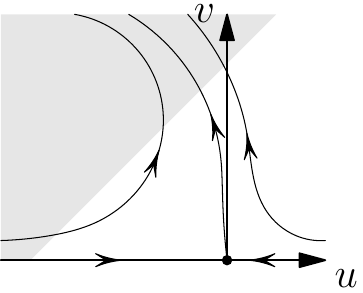}}
\caption{The landscape of free energy density $\Omega$ on the entire $uv$-plane.  The shaded region is unphysical.  The dots represent stationary points, and the arrows points toward the direction of lower $\Omega$. Only the $v > 0$ side is plotted here; the other side is symmetrical.  (a) At high temperature ($\alpha > \alpha_{c2}$), the symmetric state $u = v = 0$ is the stable minimum inside the physical region.  (b) As the temperature is lowered, the saddle point is drawn toward the symmetric state, and it is within the physical region for $\alpha_{c2} > \alpha > \alpha_{cc}$.  (We zoom in on the symmetric state, and the maximum on the far left is not shown.)  Note that the $u+2 = v$ edge may have a lower free energy than the symmetric state after the saddle point has crossed into the physical region.  (c) The saddle point eventually collapses in on the symmetric state and annihilates at $\alpha = \alpha_{cc}$.  Below this temperature there is no stable equilibrium.}
\end{figure}

Any vestigial order is indicated by $m_1 \neq m_2$.  On the chiral side $(2g_1 - g_2) < g_2$, and it is clear that for any given $\vert \vec{h} \vert$, the lowest $\Omega$ is always obtained by choosing $h_x = h_y = 0$, and the order, if exists, must be purely chiral.  We will adopt this choice from here on.

To proceed, we first introduce the dimensionless quantities:
\begin{equation}
\begin{split}
u &\equiv \left(m_1 + m_2 - 2\sqrt{a}\right)/\sqrt{a}; \\
v &\equiv \left(m_1 - m_2\right)/\sqrt{a}.
\end{split}
\label{uv}
\end{equation}
The physically meaningful range of values is $u \geq -2$, $\vert v \vert < (u+2)$.  The symmetric phase corresponds to $u = v = 0$, and a vestigial chiral order is indicated by $v \neq 0$.  The free energy \eqref{MasterFreeEnergy} has a very clean dimensionless form in terms of $u$ and $v$:
\begin{widetext}
\begin{equation}
\left(\frac{4\pi}{\alpha^{3/2}}\right) \beta (\Omega-\Omega_s)
= \left(\frac{1}{2} + \frac{(2 g_1 + g_2)}{16\pi \sqrt{\alpha}}\right) u^2 + \frac{u^3}{12}
	+ \left(\frac{1}{2} + \frac{(2g_1 - g_2)}{16\pi \sqrt{\alpha}}\right) v^2 + \frac{u v^2}{4}.
\label{ChiralFreeEnergy}
\end{equation}
\end{widetext}
This expression is bounded from below within the physical region.

The symmetric solution $u = v = 0$ is always a stationary point.  If $(2g_1 - g_2) > 0$, it is always a stable equilibrium, thus confirming our previous assertion that $(2g_1 - g_2) < 0$ is a necessary condition for a vestigial chiral phase.  When the quantity is negative, the sign of the $v^2$ coefficient changes at the instability temperature $\alpha_{cc}$:
\begin{equation}
\sqrt{\alpha_{cc}} = \frac{g_2-2g_1}{8\pi},
\end{equation}
and renders the symmetric phase unstable.  The vanishing of this coefficient is equivalent to a diverging chiral susceptibility, and may be naively taken as the critical point of a second order transition into the vestigial chiral phase.  Alas, this textbook interpretation fails utterly for the present problem: a closer look at \eqref{ChiralFreeEnergy} reveals that there exists no stable solution for $\alpha < \alpha_{cc}$.

Let us elaborate.  Since \eqref{ChiralFreeEnergy} is only a cubic polynomial, its stationary points on the full $uv$-plane can be exactly found.  In the $\alpha \rightarrow \infty$ limit, these are: local minimum $(u,v) = (0,0)$ (the symmetric solution), local maximum $(-4, 0)$, and two saddle points $(-2, \pm2)$.  Apart from $(0,0)$, the other stationary points are outside the physical region, and $(0,0)$ is the global minimum inside the physical region.  See Fig \ref{landscapeHigh} for a sketch.  

As $\alpha$ is lowered, the symmetric solution remains at $(0,0)$, the local maximum move down the $-u$-direction, and the two saddle points draw nearer toward $(0,0)$.  At $\alpha = \alpha_{c2}$ %
\footnote{
Closed form expression of $\alpha_{c2}$ can be obtained by solving the saddle point equations with the additional condition $u = v$.  It is unwieldy and not particularly useful, and we choose to omit the explicit expression here.
}
, the saddle points cross the edges $\vert v \vert = (u+2)$ into the physical region.  That is, an initially gapless ($m_2 = 0$) solution of the saddle point equations becomes available.  See Fig \ref{landscapeMid}.  This gapless solution at $\alpha_{c2}$ has special implication when we discuss the superconducting order later.  Eventually at $\alpha_{cc}$ the saddle points and the minimum merge, and leaving the symmetric solution unstable in the $v$-direction.  This is depicted in Fig \ref{landscapeLow}.

In fact, below $\alpha_{c2}$, the edges of the physical region are lower in free energy than the saddle points, and the saddle points are drawing closer to the symmetric solution continuously.  Therefore, at some point in the range $\alpha_{cc} < \alpha < \alpha_{c2}$, the edges already are lower than the symmetric solution, and \eqref{ChiralFreeEnergy} no longer has an equilibrium.  One thing left out of the analysis is the superconducting order: it is therefore natural to conjecture that a joint first order transition into the superconducting phase takes place within $\alpha_{cc} < \alpha < \alpha_{c2}$.

\subsection{Joint first order transition}

Let us now produce evidence for the above conjecture using \eqref{LWwithSSB}.  We assume a uniform chiral superconducting order $\bar{\phi}_{\up} = 0$, $\bar{\phi}_{\down} = \bar{\phi}$, chosen to be real.  There is another degenerate choice with the spin reversed.  Clearly the chiral Zeeman self energy and the chiral superconducting order mutually favor each other, and we will leave out other components in the subsequent analysis.  Given the form of the interaction \eqref{HiPhi}, one expects the pseudospin be conserved even in the presence of the superconducting order.  This leads us to write down the variational matrix propagator:
\begin{widetext}
\begin{equation}
\tilde{\mG}(p) =
\begin{pmatrix}
\ep_0(p) + h_0 + h_z & r_1 e^{-i\theta_1} & & \\
r_1 e^{i\theta_1} & \ep_0(p) + h_0 + h_z & & \\
& & \ep_0(p) + h_0 - h_z & r_2 e^{-i\theta_2} \\
& &  r_2 e^{i\theta_2} & \ep_0(p) + h_0 - h_z
\end{pmatrix}.
\end{equation}
\end{widetext}
The phases $\theta_1$ and $\theta_2$ of the off-diagonal components are undetermined \emph{a priori}, even though the superconducting order itself has been set to be real.  The amplitudes $r_1$, $r_2$ are positive.

Again, the result is most conveniently expressed in terms of the square roots of the eigenvalues of the $4\times 4$ gap matrix $\tilde{\mG}(p = 0)$.  We thus define
\begin{equation}
\begin{split}
m_{11} &= \sqrt{\alpha + h_0 + h_z + r_1},\\
m_{12} &= \sqrt{\alpha + h_0 + h_z - r_1}, \\
m_{21} &= \sqrt{\alpha + h_0 - h_z + r_2}, \\
m_{22} &= \sqrt{\alpha + h_0 - h_z - r_2}.
\end{split}
\end{equation}
It can be shown that $\theta_1$ drops out of the free energy completely, and $\theta_2 = 0$ always minimizes the free energy.  Furthermore, all stationary solutions satisfy $m_{11} - m_{12} = 0$.  So we will set $\theta_2 = 0$ and $m_{11} = m_{12} = m_1$ from here.  The resultant free energy density is
\begin{widetext}
\begin{equation}
\begin{split}
\beta\Omega &=
\frac{1}{24\pi}(2m_1^3 + m_{21}^3 + m_{22}^3) - \frac{\alpha}{8\pi}(2m_1 + m_{21} + m_{22})
+ \alpha \bar{\phi}^2 + \frac{g_1}{2} \bar{\phi}^4
\\
& \; + \frac{g_1}{128\pi^2} \left[
	8(m_1 - \sqrt{\alpha})^2 + 3(m_{21} - \sqrt{\alpha})^2 + 3(m_{22} - \sqrt{\alpha})^2
	+ 2(m_{21} - \sqrt{\alpha})^2(m_{22} - \sqrt{\alpha})^2
	\right]\\
& \; + \frac{g_2}{32\pi^2} (m_1 - \sqrt{\alpha})\left(
	m_{21}  + m_{22} - 2\sqrt{\alpha}
	\right)
 - \frac{g_1}{8\pi} \bar{\phi}^2 \left( 3 m_{21} + m_{22} - 4\sqrt{\alpha} \right)
- \frac{g_2}{4\pi} \bar{\phi}^2 \left( m_1 - \sqrt{\alpha} \right).
\end{split}
\label{ChiralWithSC}
\end{equation}
\end{widetext}

\begin{figure}
\subfloat[\label{phiBarPlot}]{\includegraphics[width=0.4\textwidth]{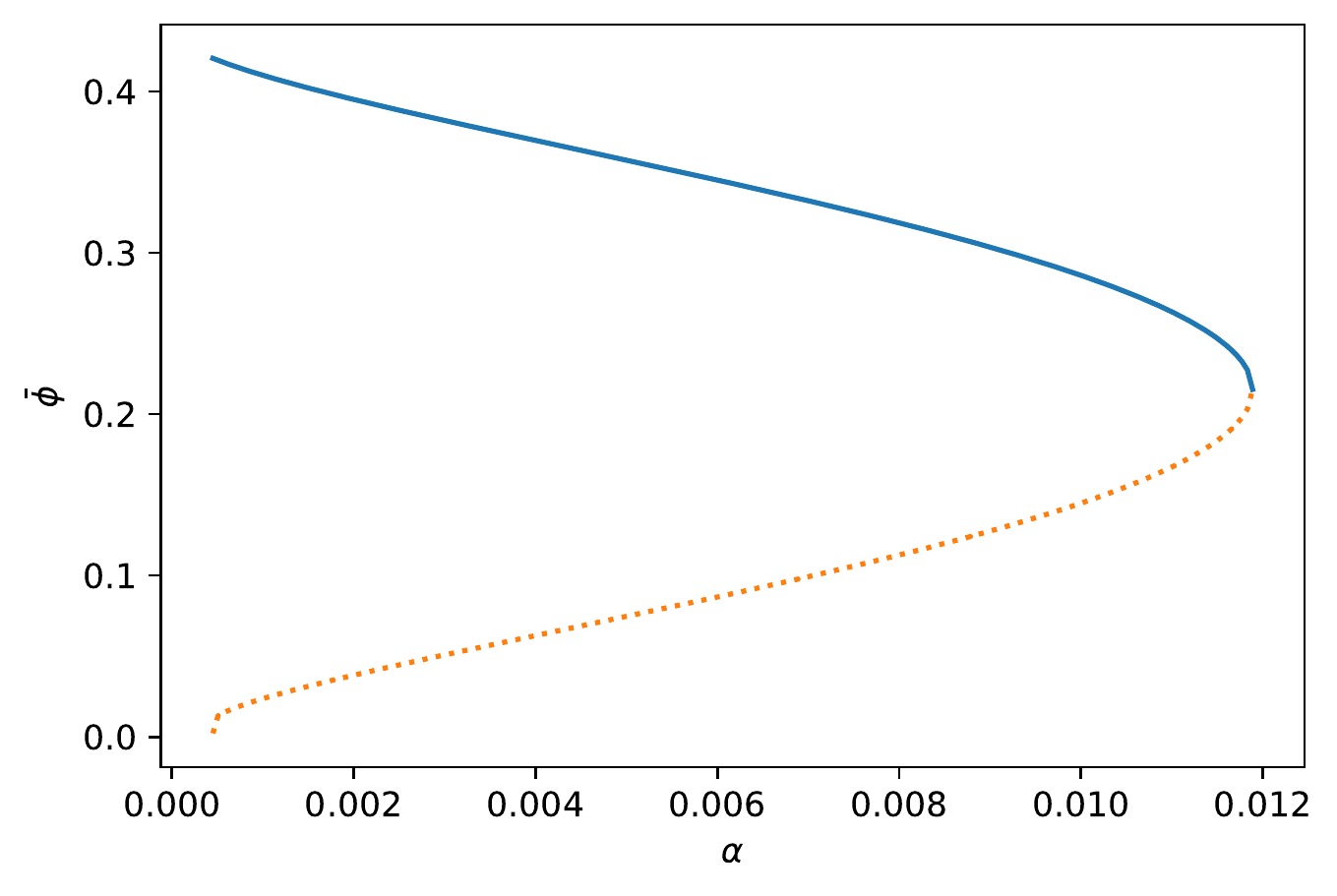}} \qquad
\subfloat[\label{m22Plot}]{\includegraphics[width=0.24\textwidth]{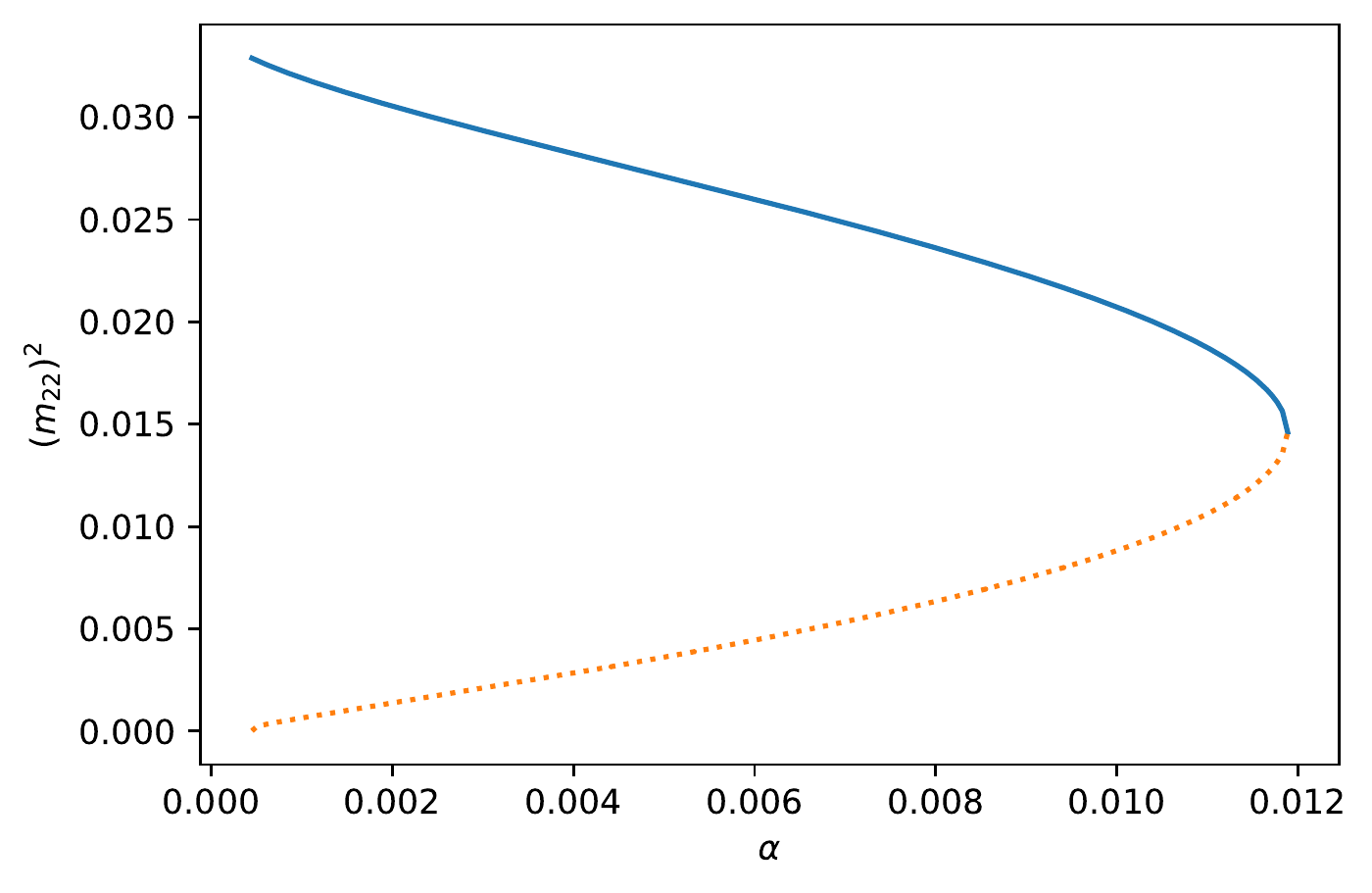}}
\;
\subfloat[\label{artifactPlot}]{\includegraphics[width=0.2\textwidth]{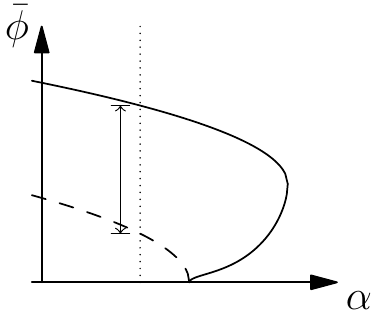}}
\caption{(a) Superconducting order parameter $\bar{\phi}$ and (b) smallest energy gap $m_{22}^2$ versus $\alpha$ for the superconducting solution, plotted at $g_2/g_1 = 2.5$.  The dotted branch is the unstable solution that annihilates as it becomes gapless at $\alpha_{cc}$ slightly bigger than zero, and the solid line is the (meta-)stable branch that continues onto lower temperature. $\bar{\phi}$ is in unit of $\sqrt{g_1}$, and everything else in unit of $g_1^2$.  We see that $m_{22}^2 \neq 0$ explicitly violates Goldstone's theorem. (c) The schematic illustration of the conjectured actual superconducting solution (dashed) alongside the Hartree-Fock solution (solid).  The dotted line represents the first order transition temperature.  As marked on the diagram, the artifact dominates the apparent first order jump given by the Hartree-Fock solution.}
\end{figure}

We proceed to solve the saddle point equations $\partial\Omega/\partial m_1 = \partial\Omega/\partial m_{21} = \partial\Omega/\partial m_{22} = \partial\Omega/\partial \bar{\phi} = 0$.  All previously found stationary points (the symmetric solution and the pair of saddle points) remain solutions with $\bar{\phi} = 0$.  In addition, there exist stationary points with $\bar{\phi} \neq 0$ not connected to the symmetric solution.  As the temperature is lowered, a pair of $\bar{\phi} \neq 0$ solutions come into existence through a saddle-node bifurcation, one local maximum and the other minimum.  See Fig \ref{phiBarPlot}.  When the temperature is further lowered, the maximum is continuously connected to the aforementioned gapless, non-superconducting solution at $\alpha_{c2}$, and ceases to exist for $\alpha < \alpha_{c2}$.  The other solution rapidly overtakes the symmetric state to become the global minimum in free energy at some $\alpha > \alpha_{cc}$, before the symmetric state turns unstable.  Neither branch of the superconducting solutions exhibit a gapless Goldstone mode: see Fig 
\ref{m22Plot}.

This is reminiscent to the typical result obtained from similar self-energy methods; see ref \citep{Baym1977-lm} and in particular appendix G of \cite{Hugel2016-bx}.  In these other reported cases, the bifurcation structure is seen as an artifact of the method: the normal gapless solution is physically expected to be the \emph{onset of degenerate superconducting minima}, rather than the termination of degenerate maxima.  We therefore conjecture the following: a pair superconducting local minima (related by time-reversal) grow out of the gapless, normal solution at $\alpha_{c2}$.  They continue to exist and becomes the true equilibrium solution at some $\alpha > \alpha_{c}$.  See Fig \ref{artifactPlot}.

Apart from qualitatively establishing the joint first order transition into the superconducting phase, we believe that any quantitative results here should be taken with more than a few grains of salt.  In particular, \eqref{ChiralWithSC} indicates a very strong first order transition, but we believe much of this jump is artifact of the self energy method.  Unfortunately we have no other way to estimate the magnitude of the jump.

\subsection{Away from the decoupled limit}

Finally, one may restore $C_1, C_2 \neq 0$ and re-do the calculation to see if the effective spin-orbit coupling changes the picture.  The renormalization counter term $\delta\alpha$ is changed accordingly:
\begin{equation}
\delta\alpha = -(2g_1 + g_2)\int_k \frac{\ep_0(k)}{\ep_0(k)^2 - \ep_x(k)^2 - \ep_y(k)^2}.
\end{equation}
An analytic evaluation of $\Omega$ is now impossible even in the non-interacting limit, and we perform an expansion in powers of $C_1$ and $C_2$.  It can be seen that the first order correction vanishes, and quadratic correction terms are proportional to the parameter $J \equiv (\frac{8}{15} C_1^2 + \frac{2}{15} C_2^2)$, coming from the angular average of $(\ep_x^2 + \ep_y^2)$.  We estimate the typical value of $J$ would be of the order $0.1$ or smaller (see apendix), though we cannot definitely rule out a bigger, more extreme value.  As noted in \cite{Hecker2018-bb}, the theory is still cylindrically symmetric at this order; the three-fold anisotropy sets in only at cubic order in $C_1$ and $C_2$.  In the absence of a superconducting order, this first correction to the free energy reads:
\begin{widetext}
\begin{equation}
\begin{split}
\beta \Omega &= (\text{decoupled limit}) \\
&\; + \frac{J}{8\pi}  (m_1 + m_2)^{-1} \, \Bigg \lbrace \left[(m_1^4 + m_1^3m_2 + m_1^2m_2^2 + m_1m_2^3 + m_2^4) - 5\alpha(m_1^2 + m_1 m_2 + m_2^2)\right] \\
&\qquad + \frac{(2g_1+g_2)}{8\pi} \left[
	5(m_1^3 + 2m_1^2m_2+2m_1m_2^2+m_2^3)
	- \frac{5}{2}\sqrt{\alpha}(5m_1^2+14m_1m_2+5m_2^2)\right] \\
&\qquad + \frac{(2g_1-g_2)}{8\pi} \left[
	\frac{5}{4} (m_1 + m_2)(m_1 - m_2)^2
	- \frac{(m_1-m_2)^4}{4(m_1+m_2)}\right] \Bigg \rbrace + \dots
\end{split}
\label{ChiralFreeEnergyAnisotropic1}
\end{equation}
$m_1$ and $m_2$ are defined in \eqref{GapEigenvalues}.
\end{widetext}

We adopt a different strategy to analyze this free energy.  The above expression is Taylor expanded in terms of the dimensionless variables $u$ and $v$ introduced in \eqref{uv}, and the series is truncated to contain only $u^2$, $u v^2$, $v^2$, and $v^4$, in the spirit of the GL theory.  Completing the square on $u$ generates a (negative) correction to the $v^4$ coefficient, valid for $v$ small enough.  If the overall effective quartic coefficient is positive, we argue that a second order transition to the vestigial phase does take place; otherwise the joint first order transition remains the only possibility.  Applying the criterion to the decoupled limit \eqref{ChiralFreeEnergy}, one sees that the intrinsic $v^4$ coefficient is absent, and the correction from $u$-fluctuation makes the overall effective $v^4$ coefficient negative.  Indeed there is no vestigial phase as we have already concluded.  For the present expression \eqref{ChiralFreeEnergyAnisotropic1}, we have
\begin{equation}
\begin{split}
\left(\frac{4\pi}{\alpha^{3/2}}\right)\beta\Omega
&\approx
\left[\frac{1}{2} + \frac{15}{16}J + \left(\frac{2g_1+g_2}{16\pi\sqrt{\alpha}}\right)
			\left(1+\frac{15}{4} J\right)\right] u^2 \\
&\quad + \left[\frac{1}{2} + \frac{5}{16}J + \left(\frac{2g_1-g_2}{16\pi\sqrt{\alpha}}\right)
			\left(1+\frac{5}{4}J\right)\right] v^2 \\
&\quad + \left[\frac{1}{4} + \frac{15}{32}J + \left(\frac{2g_1+g_2}{16\pi\sqrt{\alpha}}\right)
			\frac{5}{8}J\right] uv^2\\
&\quad + J \left[\frac{1}{64} - \left(\frac{2g_1-g_2}{16\pi\sqrt{\alpha}}\right) \frac{1}{16}\right] v^4.
\end{split}
\end{equation}

At the first glance, the effective spin-orbit coupling does tend to stabilize the vestigial phase by generating a positive $v^4$ term.  $(2g_1-g_2) < 0$ is still required for non-trivial behavior, and the instability temperature $\alpha_{cc}$ now takes an $O(J)$ correction:
\begin{equation}
\sqrt{\alpha_{cc}} = \frac{(g_2-2g_1)}{8\pi} \left(1 + \frac{5}{8} J + \dots\right).
\end{equation}

In principle one would want to expand every expression to $O(J)$, but the ratio $R_c \equiv (g_2 + 2g_1)/(g_2-2g_1) \geq 1$ is unconstrained and potentially very big.  The product $R_c J \geq J$ may or may not be small, and extra attention is due when analyzing the result.  At $\alpha_{cc}$, taking into account the $u$-fluctuation, the effective $v^4$ coefficient is
\begin{equation}
\left(\frac{3}{64}J + O(J^2)\right)
	-\frac{ 
		\left[
		\frac{1}{4} + \frac{15}{32} J + \left(\frac{5J}{16+10J}\right) R_c
		\right]^2
		}
		{
		\frac{1}{3} + \frac{15}{16}J + \left( \frac{4 + 15J}{8+5J} \right) R_c
		}.
\label{chiralV4}
\end{equation}
The first term is the ``intrinsic'' $v^4$ coefficient, and the second term comes from the $u$-fluctuation.  It can be numerically checked that \eqref{chiralV4} is never positive for any $R_c \geq 1$ for any given $J$.  Nevertheless, the expression is only meaningful for small $J$, and we can only rule out the vestigial chiral phase when effective spin-orbit coupling is not strong.

In conclusion, for a range of physically reasonable parameters, \emph{the vestigial chiral order cannot exist} above a weak-coupling two-component chiral superconductor via the Ginzburg-Landau mechanism investigated in the present paper and \cite{Hecker2018-bb, Fernandes2019-yv}.  Even though there is an apparent divergent in chiral susceptibility, this does not indicate a second order phase transition, since there is no (meta)-stable non-superconducting state available below the instability temperature $\alpha_{cc}$.  Instead, the system undergoes a joint first order transition into the superconducting phase at a temperature slightly above $\alpha_{cc}$.  This negative result holds for arbitrary ratio of coupling constants $R_c$, but requires the coefficient $J$ to be small.  Recall that $J$ is a measure of the size of anisotropic gradient terms in the effective Hamiltonian.  The typical value of $J$ is small, however, and it already takes some extreme choice of parameters to bring $J \sim O(1)$.

\subsection{Comparison with Fischer and Berg}

Fischer and Berg \cite{Fischer2016-gt} considered a very similar model of a two-component order parameters with tetragonal symmetry, instead of the trigonal symmetry considered in this work.  They identified effectively the same $g_2 > 2g_1$ criterion.  They also found a stable vestigial chiral phase at larger $g_2$ in their numerical calculation.

Technically, our model and theirs differ in two ways.  First, using the second form of \eqref{HiPhi}, all three $\lambda_{x,y,z}$ can take arbitrary values in the tetragonal case, while the trigonal symmetry imposes $\lambda_x = \lambda_y$ in addition.  Second, the form of allowed effective spin-orbit coupling is different under tetragonal symmetry.  To the order at which we are working, the spin-orbit coupling term is replaced by its angular average, and the difference amounts to a change in the $J$ parameter.  And the interaction terms practically have the same form with different coupling constants.  Thus our negative result, with minimal modification, should be applicable for the tetragonal symmetry, too.

One key technical difference between our method and Fischer and Berg's is the removal of UV-divergence.  We argue that (for a weak-coupling superconductor) the long-wavelength physics of the finite-temperature phase transition cannot depend on anything at the atomic scale, that all relevant scales of the problem are much smaller compared to the size of the Brillouin zone, and it is very natural to subtract off the UV-divergence.  And we expect to see small $g_1$ and $g_2$ compared with lattice scales, reflecting the smallness of the Ginzburg parameter or the ratio $T_c/E_F$.

On the other hand, Fischer and Berg regularized their model with a lattice.  The inverse lattice spacing becomes their momentum unit, and $T_c$ is their energy unit.  The vestigial chiral phase found in their numerical calculation required not only a large ratio of $(g_2 - 2g_1)/(g_2 + 2g_1)$, but also that $g_1$, $g_2$ individually of order unity in their chosen units.  It is not clear to us that their theory contained any small parameter, and the parameter range where the vestigial chiral phase was found seems unlikely to be realized in real materials.  They did not find a vestigial chiral phase when the coupling constant is sufficiently small compared with the cutoff; this is consistent with our negative result.

We also note that the harmonic variational approach adopted by Fischer and Berg is in effect identical to our LW approach with Hartree-Fock approximation.  The LW approach, however, offers a clear direction for systematic improvement.

\section{The nematic side}

\subsection{Non-existence of vestigial order}

Most of the result from the previous section can be directly transplanted to the nematic case ($g_2 < g_1$).  The free energy \eqref{MasterFreeEnergy} is valid irrespective to the values of $g_1$ and $g_2$.  This time, one always chooses $h_z = 0$ for the minimum, and any possible vestigial order is always purely nematic.  We note that the theory in the decoupled limit does not possess a preferred nematic orientation due to its full cylindrical symmetry; an orientation must be randomly and spontaneously chosen in the ordered phase.  This is but an artifact of the forth order GL theory in the decoupled limit: the effective spin-orbit terms and the sixth order terms in the GL expansion would breaks the rotational symmetry down to six-fold \cite{Venderbos2016-fa, How2019-re}.

Once $h_z$ is set to zero, one recognizes that all results in section \ref{4a} automatically apply if $(2g_1 - g_2)$ is replaced by $g_2$.  In particular, the symmetric solution becomes unstable if $g_2 < 0$, and the instability occurs at $\alpha_{cn} = (g_2/8\pi)^2$.  But no vestigial nematic phase is to be found, as no stable minimum of free energy exists below $\alpha_{cn}$ without superconducting order.  A joint first order superconducting transition at a temperature slightly above $\alpha_{cn}$ is again the real answer.

The calculation for correction due to effective spin-orbit coupling proceeds in the same manner, but the coefficients naturally turn out to have different values.  The leading correction to the free energy is
\begin{widetext}
\begin{equation}
\begin{split}
\beta \Omega &= (\text{decoupled limit}) \\
&\; + \frac{J}{64\pi}  (m_1 + m_2)^{-1} \, \Bigg \lbrace \left[(9m_1^4 + 9m_1^3m_2 + 4m_1^2m_2^2 + 9 m_1m_2^3 + 9m_2^4) - 5\alpha(7m_1^2 + 10m_1 m_2 + 7m_2^2)\right] \\
&\quad + \frac{(2g_1+g_2)}{8\pi} \left[
	5(7m_1^3 + 17m_1^2m_2 + 17m_1m_2^2 + 7m_2^3)
	- 10\sqrt{\alpha}(13m_1^2 + 22m_1m_2 + 13m_2^2)\right] \\
&\quad + \frac{g_2}{8\pi} \left[
	19 (m_1 + m_2)(m_1 - m_2)^2
	+ \frac{4m_1m_2(m_1-m_2)^2}{(m_1+m_2)}\right] \Bigg \rbrace + \dots
\end{split}
\end{equation}
\end{widetext}

The instability temperature is again shifted by $O(J)$:
\begin{equation}
\sqrt{\alpha_{cn}} = -\frac{g_2}{8\pi}\left( 1 + \frac{5}{4} J + \dots\right).
\end{equation}
Expanding in terms of $u, v$ defined in \eqref{uv} yields
\begin{equation}
\begin{split}
\left(\frac{4\pi}{\alpha^{3/2}}\right)\beta\Omega
&\approx
\left[\frac{1}{2} + \frac{15}{16}J + \left(\frac{2g_1+g_2}{16\pi\sqrt{\alpha}}\right)
			\left(1+\frac{15}{4}J\right)\right] u^2 \\
&\quad + \left[\frac{1}{2} + \frac{5}{8}J + \left(\frac{g_2}{16\pi\sqrt{\alpha}}\right)
			\left(1+\frac{5}{2}J\right)\right] v^2 \\
&\quad + \left[\frac{1}{4} + \frac{15}{32}J + \left(\frac{2g_1+g_2}{16\pi\sqrt{\alpha}}\right)
			\frac{5}{16}J\right] uv^2\\
&\quad + J \left[\frac{1}{128} - \left(\frac{g_2}{16\pi\sqrt{\alpha}}\right) \frac{1}{32}\right] v^4.
\end{split}
\end{equation}
Going through the same exercise of eliminating $u$ using its saddle point equation, the overall effective $v^4$ coefficient thus generated is again found to be always negative for all $J$.  (Again, note that the approximation is not valid for large-$J$.)  We can once again conclude that \emph{there is no vestigial nematic phase when $J$ is sufficiently small to justify the perturbative treatment}.

\subsection{Symmetry and renormalization group}

\begin{figure}
\includegraphics[scale=1]{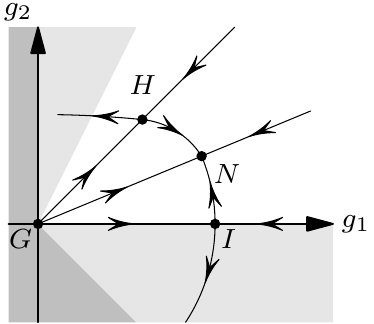}
\caption{\label{RG} The perturbative RG flow.  The dark shaded region is unstable, and the light shaded regions are the deep nematic ($g_2 < 0$) and deep chiral ($g_2 > 2g_1$) regions.  The four fixed points are labeled on the diagram: G for Gaussian, H for Heisenberg, I for Ising, and N for nematic.}
\end{figure}

There is in fact a lot of prior hints on our negative result.  The renormalization group analysis \cite{Patashinskii1979-hr} in $d=4-\ep$ reveals four fixed points for our model: the Gaussian fixed point $g_1 = g_2 = 0$, the Heisenberg fixed point with $g_1 = g_2$, the Ising fixed point with $g_2 = 0$, and the nematic fixed point in between; see Fig \ref{RG}.  Without effective spin-orbit coupling, at one-loop order, the nematic fixed point sits on top the Heisenberg fixed point; the separation of the two is stabilized either by the two-loop contribution or by the inclusion of spin-orbit coupling as a perturbation.  This suggests the picture: for $g_1 > g_2 > 0$, the system goes from the normal symmetric phase straight into the nematic superconducting phase through a second order transition, and something else must happen for $g_2 < 0$.

Along the $g_2 = 0$ line, the two pseudospins are not coupled by the interaction.  If we further switch off the effective spin-orbit coupling, the theory enjoys an enhanced $SU(2)\times SU(2)$ symmetry.  The $g_2 = 0$ boundary should therefore be very robust given the symmetry protection.  And it also makes ample sense that a line of enhanced symmetry separates the two regions in the parameter space that have markedly different critical behaviors.

The fact that vestigial order is not to be found in the $g_2 < 0$ region may be more surprising.  However, we note that conventional wisdom \cite{Patashinskii1979-hr} indeed calls for a first order transition for this and other similar situations.

It is much harder to frame the story on the chiral side in terms of renormalization group.  Perurbative calculation reveals no fixed point for $g_2 > g_1$.  While there exist claims of a chiral fixed point \cite{Kawamura1986-mo, Kawamura1988-ga}, the result appears to at least require a number of spin components much larger than two \cite{Antonenko1995-ix, Delamotte2004-rv, Sorokin2019-ed}.  And more importantly, nothing special can be said about the $g_2 = 2g_1$ line.

\section{Conclusion}

The central (negative) result presented in this paper is the following: \emph{within Ginzburg-Landau theory with fluctuations, in a region of the parameter space centered around the effective spin-orbit-decoupled limit, vestigial order cannot exist in the normal state of a weak-coupling two-component superconductor.}  We argue in the appendix that the physically relevant range of parameters is near the decoupled limit.

\begin{figure}
\includegraphics[width=0.4\textwidth]{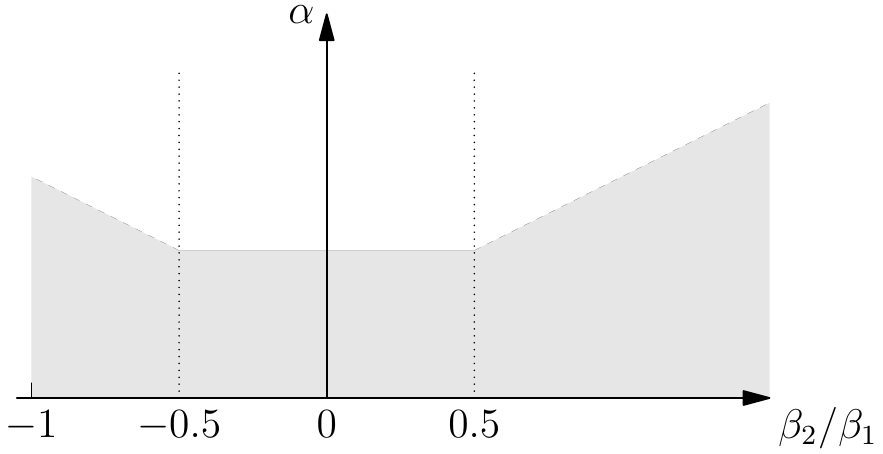}
\caption{\label{phaseDiagram} The proposed phase diagram.  The shaded region represents superconductivity, and the unshaded, high-temperature phase is symmetric (normal without vestigial order).  The solid phase boundary represents second order transition, and the dashed boundary represents joint first order transition.}
\end{figure}

To recap, for $0 < g_2 < 2g_1$ we still obtain a second order phase transition into the nematic or chiral superconducting phase, similar to the simple meanfield prediction.  The system must be outside this range, i.e. in the respective deep nematic or chiral regime, to deviate from the meanfield behavior.  And even then, we predict a joint first order transition rather than a vestigial order phase.  The $0 < g_2 < 2g_1$ criterion was priorly obtained by Fischer and Berg \cite{Fischer2016-gt}.  This is summarized in the phase diagram Fig \ref{phaseDiagram}.

On the nematic side, existing renormalization group analysis and symmetry argument \cite{Patashinskii1979-hr} already hinted at the $g_2 < 0$ criterion.  But it does comes as a surprise that no vestigial nematic phase is found at all.  Our result contradicts Hecker and Schmalian's \cite{Hecker2018-bb}.  We believe the Hubbard-Stratonovich decoupling adopted in their analysis underestimates the interaction strength and results in their error.  It is also worth noting that the existing calculations of GL coefficients from various microscopic models (see supplementary materials of \cite{Venderbos2016-cv, Zyuzin2017-kn} for the calculation; also see footnote [39]) all results in $g_2 > 0$; that is, none is in the deep nematic region.

Fischer and Berg \cite{Fischer2016-gt} found a stable vestigial chiral phase for a similar model with tetragonal symmetry, but only when the coupling constants are comparable in size to the large-momentum cutoff in their numerical analysis.  We argue that this regime is not physical relevant.  For weaker coupling, their result is consistent with ours.

One certain conclusion can be drawn from the this work: the normal state anisotropy observed in $M_x$Bi$_2$Se$_3$ \cite{Sun2019-kt, Cho2020-ne} cannot be interpreted as a vestigial order associated to the underlying superconductivity.  Whatever else the origin of this normal state anisotropy is, it is likely to have important implication on the pairing mechanism leading to the superconductivity.  Our result can readily be generalized to other unconventional superconductors and superfluids with two-component order parameters but different symmetries.

\begin{acknowledgments}
This work is supported by National Science and Technology Council (NSTC), Taiwan under grant No. MOST-110-2112-M-001-051-MY3.  P.T.H. is supported under Grant No. MOST 110-2811-M-001-561.
\end{acknowledgments}

\bibliography{ref}

\begin{thebibliography}{63}%
\makeatletter
\providecommand \@ifxundefined [1]{%
 \@ifx{#1\undefined}
}%
\providecommand \@ifnum [1]{%
 \ifnum #1\expandafter \@firstoftwo
 \else \expandafter \@secondoftwo
 \fi
}%
\providecommand \@ifx [1]{%
 \ifx #1\expandafter \@firstoftwo
 \else \expandafter \@secondoftwo
 \fi
}%
\providecommand \natexlab [1]{#1}%
\providecommand \enquote  [1]{``#1''}%
\providecommand \bibnamefont  [1]{#1}%
\providecommand \bibfnamefont [1]{#1}%
\providecommand \citenamefont [1]{#1}%
\providecommand \href@noop [0]{\@secondoftwo}%
\providecommand \href [0]{\begingroup \@sanitize@url \@href}%
\providecommand \@href[1]{\@@startlink{#1}\@@href}%
\providecommand \@@href[1]{\endgroup#1\@@endlink}%
\providecommand \@sanitize@url [0]{\catcode `\\12\catcode `\$12\catcode
  `\&12\catcode `\#12\catcode `\^12\catcode `\_12\catcode `\%12\relax}%
\providecommand \@@startlink[1]{}%
\providecommand \@@endlink[0]{}%
\providecommand \url  [0]{\begingroup\@sanitize@url \@url }%
\providecommand \@url [1]{\endgroup\@href {#1}{\urlprefix }}%
\providecommand \urlprefix  [0]{URL }%
\providecommand \Eprint [0]{\href }%
\providecommand \doibase [0]{https://doi.org/}%
\providecommand \selectlanguage [0]{\@gobble}%
\providecommand \bibinfo  [0]{\@secondoftwo}%
\providecommand \bibfield  [0]{\@secondoftwo}%
\providecommand \translation [1]{[#1]}%
\providecommand \BibitemOpen [0]{}%
\providecommand \bibitemStop [0]{}%
\providecommand \bibitemNoStop [0]{.\EOS\space}%
\providecommand \EOS [0]{\spacefactor3000\relax}%
\providecommand \BibitemShut  [1]{\csname bibitem#1\endcsname}%
\let\auto@bib@innerbib\@empty
\bibitem [{\citenamefont {Matano}\ \emph {et~al.}(2016)\citenamefont {Matano},
  \citenamefont {Kriener}, \citenamefont {Segawa}, \citenamefont {Ando},\ and\
  \citenamefont {Zheng}}]{Matano2016-vr}%
  \BibitemOpen
  \bibfield  {author} {\bibinfo {author} {\bibfnamefont {K.}~\bibnamefont
  {Matano}}, \bibinfo {author} {\bibfnamefont {M.}~\bibnamefont {Kriener}},
  \bibinfo {author} {\bibfnamefont {K.}~\bibnamefont {Segawa}}, \bibinfo
  {author} {\bibfnamefont {Y.}~\bibnamefont {Ando}},\ and\ \bibinfo {author}
  {\bibfnamefont {G.-Q.}\ \bibnamefont {Zheng}},\ }\bibfield  {title} {\bibinfo
  {title} {Spin-rotation symmetry breaking in the superconducting state of cu x
  {Bi2Se3}},\ }\href@noop {} {\bibfield  {journal} {\bibinfo  {journal} {Nat.
  Phys.}\ }\textbf {\bibinfo {volume} {12}},\ \bibinfo {pages} {852} (\bibinfo
  {year} {2016})}\BibitemShut {NoStop}%
\bibitem [{\citenamefont {Pan}\ \emph {et~al.}(2016)\citenamefont {Pan},
  \citenamefont {Nikitin}, \citenamefont {Araizi}, \citenamefont {Huang},
  \citenamefont {Matsushita}, \citenamefont {Naka},\ and\ \citenamefont
  {de~Visser}}]{Pan2016-ni}%
  \BibitemOpen
  \bibfield  {author} {\bibinfo {author} {\bibfnamefont {Y.}~\bibnamefont
  {Pan}}, \bibinfo {author} {\bibfnamefont {A.~M.}\ \bibnamefont {Nikitin}},
  \bibinfo {author} {\bibfnamefont {G.~K.}\ \bibnamefont {Araizi}}, \bibinfo
  {author} {\bibfnamefont {Y.~K.}\ \bibnamefont {Huang}}, \bibinfo {author}
  {\bibfnamefont {Y.}~\bibnamefont {Matsushita}}, \bibinfo {author}
  {\bibfnamefont {T.}~\bibnamefont {Naka}},\ and\ \bibinfo {author}
  {\bibfnamefont {A.}~\bibnamefont {de~Visser}},\ }\bibfield  {title} {\bibinfo
  {title} {Rotational symmetry breaking in the topological superconductor sr x
  {Bi2Se3} probed by upper-critical field experiments},\ }\href@noop {}
  {\bibfield  {journal} {\bibinfo  {journal} {Sci. Rep.}\ }\textbf {\bibinfo
  {volume} {6}},\ \bibinfo {pages} {28632} (\bibinfo {year}
  {2016})}\BibitemShut {NoStop}%
\bibitem [{\citenamefont {Asaba}\ \emph {et~al.}(2017)\citenamefont {Asaba},
  \citenamefont {Lawson}, \citenamefont {Tinsman}, \citenamefont {Chen},
  \citenamefont {Corbae}, \citenamefont {Li}, \citenamefont {Qiu},
  \citenamefont {Hor}, \citenamefont {Fu},\ and\ \citenamefont
  {Li}}]{Asaba2017-vn}%
  \BibitemOpen
  \bibfield  {author} {\bibinfo {author} {\bibfnamefont {T.}~\bibnamefont
  {Asaba}}, \bibinfo {author} {\bibfnamefont {B.~J.}\ \bibnamefont {Lawson}},
  \bibinfo {author} {\bibfnamefont {C.}~\bibnamefont {Tinsman}}, \bibinfo
  {author} {\bibfnamefont {L.}~\bibnamefont {Chen}}, \bibinfo {author}
  {\bibfnamefont {P.}~\bibnamefont {Corbae}}, \bibinfo {author} {\bibfnamefont
  {G.}~\bibnamefont {Li}}, \bibinfo {author} {\bibfnamefont {Y.}~\bibnamefont
  {Qiu}}, \bibinfo {author} {\bibfnamefont {Y.~S.}\ \bibnamefont {Hor}},
  \bibinfo {author} {\bibfnamefont {L.}~\bibnamefont {Fu}},\ and\ \bibinfo
  {author} {\bibfnamefont {L.}~\bibnamefont {Li}},\ }\bibfield  {title}
  {\bibinfo {title} {Rotational symmetry breaking in a trigonal superconductor
  nb-doped {Bi2Se3}},\ }\href@noop {} {\bibfield  {journal} {\bibinfo
  {journal} {Physical Review X}\ }\textbf {\bibinfo {volume} {7}},\ \bibinfo
  {pages} {011009} (\bibinfo {year} {2017})}\BibitemShut {NoStop}%
\bibitem [{\citenamefont {Shen}\ \emph {et~al.}(2017)\citenamefont {Shen},
  \citenamefont {He}, \citenamefont {Yuan}, \citenamefont {Huang},
  \citenamefont {Cho}, \citenamefont {Lee}, \citenamefont {Hor}, \citenamefont
  {Law},\ and\ \citenamefont {Lortz}}]{Shen2017-bq}%
  \BibitemOpen
  \bibfield  {author} {\bibinfo {author} {\bibfnamefont {J.}~\bibnamefont
  {Shen}}, \bibinfo {author} {\bibfnamefont {W.-Y.}\ \bibnamefont {He}},
  \bibinfo {author} {\bibfnamefont {N.~F.~Q.}\ \bibnamefont {Yuan}}, \bibinfo
  {author} {\bibfnamefont {Z.}~\bibnamefont {Huang}}, \bibinfo {author}
  {\bibfnamefont {C.-W.}\ \bibnamefont {Cho}}, \bibinfo {author} {\bibfnamefont
  {S.~H.}\ \bibnamefont {Lee}}, \bibinfo {author} {\bibfnamefont {Y.~S.}\
  \bibnamefont {Hor}}, \bibinfo {author} {\bibfnamefont {K.~T.}\ \bibnamefont
  {Law}},\ and\ \bibinfo {author} {\bibfnamefont {R.}~\bibnamefont {Lortz}},\
  }\bibfield  {title} {\bibinfo {title} {Nematic topological superconducting
  phase in nb-doped {Bi2Se3}},\ }\href@noop {} {\bibfield  {journal} {\bibinfo
  {journal} {npj Quantum Materials}\ }\textbf {\bibinfo {volume} {2}},\
  \bibinfo {pages} {59} (\bibinfo {year} {2017})}\BibitemShut {NoStop}%
\bibitem [{\citenamefont {Yonezawa}\ \emph {et~al.}(2017)\citenamefont
  {Yonezawa}, \citenamefont {Tajiri}, \citenamefont {Nakata}, \citenamefont
  {Nagai}, \citenamefont {Wang}, \citenamefont {Segawa}, \citenamefont {Ando},\
  and\ \citenamefont {Maeno}}]{Yonezawa2017-fr}%
  \BibitemOpen
  \bibfield  {author} {\bibinfo {author} {\bibfnamefont {S.}~\bibnamefont
  {Yonezawa}}, \bibinfo {author} {\bibfnamefont {K.}~\bibnamefont {Tajiri}},
  \bibinfo {author} {\bibfnamefont {S.}~\bibnamefont {Nakata}}, \bibinfo
  {author} {\bibfnamefont {Y.}~\bibnamefont {Nagai}}, \bibinfo {author}
  {\bibfnamefont {Z.}~\bibnamefont {Wang}}, \bibinfo {author} {\bibfnamefont
  {K.}~\bibnamefont {Segawa}}, \bibinfo {author} {\bibfnamefont
  {Y.}~\bibnamefont {Ando}},\ and\ \bibinfo {author} {\bibfnamefont
  {Y.}~\bibnamefont {Maeno}},\ }\bibfield  {title} {\bibinfo {title}
  {Thermodynamic evidence for nematic superconductivity in cu x {Bi2Se3}},\
  }\href@noop {} {\bibfield  {journal} {\bibinfo  {journal} {Nat. Phys.}\
  }\textbf {\bibinfo {volume} {13}},\ \bibinfo {pages} {123} (\bibinfo {year}
  {2017})}\BibitemShut {NoStop}%
\bibitem [{\citenamefont {Smylie}\ \emph {et~al.}(2018)\citenamefont {Smylie},
  \citenamefont {Willa}, \citenamefont {Claus}, \citenamefont {Koshelev},
  \citenamefont {Song}, \citenamefont {Kwok}, \citenamefont {Islam},
  \citenamefont {Gu}, \citenamefont {Schneeloch}, \citenamefont {Zhong},\ and\
  \citenamefont {Welp}}]{Smylie2018-ab}%
  \BibitemOpen
  \bibfield  {author} {\bibinfo {author} {\bibfnamefont {M.~P.}\ \bibnamefont
  {Smylie}}, \bibinfo {author} {\bibfnamefont {K.}~\bibnamefont {Willa}},
  \bibinfo {author} {\bibfnamefont {H.}~\bibnamefont {Claus}}, \bibinfo
  {author} {\bibfnamefont {A.~E.}\ \bibnamefont {Koshelev}}, \bibinfo {author}
  {\bibfnamefont {K.~W.}\ \bibnamefont {Song}}, \bibinfo {author}
  {\bibfnamefont {W.-K.}\ \bibnamefont {Kwok}}, \bibinfo {author}
  {\bibfnamefont {Z.}~\bibnamefont {Islam}}, \bibinfo {author} {\bibfnamefont
  {G.~D.}\ \bibnamefont {Gu}}, \bibinfo {author} {\bibfnamefont {J.~A.}\
  \bibnamefont {Schneeloch}}, \bibinfo {author} {\bibfnamefont {R.~D.}\
  \bibnamefont {Zhong}},\ and\ \bibinfo {author} {\bibfnamefont
  {U.}~\bibnamefont {Welp}},\ }\bibfield  {title} {\bibinfo {title}
  {Superconducting and normal-state anisotropy of the doped topological
  insulator {Sr0.1Bi2Se3}},\ }\href@noop {} {\bibfield  {journal} {\bibinfo
  {journal} {Sci. Rep.}\ }\textbf {\bibinfo {volume} {8}},\ \bibinfo {pages}
  {7666} (\bibinfo {year} {2018})}\BibitemShut {NoStop}%
\bibitem [{\citenamefont {Yonezawa}(2018)}]{Yonezawa2018-to}%
  \BibitemOpen
  \bibfield  {author} {\bibinfo {author} {\bibfnamefont {S.}~\bibnamefont
  {Yonezawa}},\ }\bibfield  {title} {\bibinfo {title} {Nematic
  superconductivity in doped {Bi2Se3} topological superconductors},\
  }\href@noop {} {\bibfield  {journal} {\bibinfo  {journal} {Z. Phys. B:
  Condens. Matter}\ }\textbf {\bibinfo {volume} {4}},\ \bibinfo {pages} {2}
  (\bibinfo {year} {2018})}\BibitemShut {NoStop}%
\bibitem [{\citenamefont {Fu}(2014)}]{Fu2014-gm}%
  \BibitemOpen
  \bibfield  {author} {\bibinfo {author} {\bibfnamefont {L.}~\bibnamefont
  {Fu}},\ }\bibfield  {title} {\bibinfo {title} {Odd-parity topological
  superconductor with nematic order: Application to
  {Cu$_{x}$Bi$_{2}$Se$_{3}$}},\ }\href@noop {} {\bibfield  {journal} {\bibinfo
  {journal} {Phys. Rev. B: Condens. Matter Mater. Phys.}\ }\textbf {\bibinfo
  {volume} {90}},\ \bibinfo {pages} {100509} (\bibinfo {year}
  {2014})}\BibitemShut {NoStop}%
\bibitem [{\citenamefont {Sauls}(1994)}]{Sauls1994-ol}%
  \BibitemOpen
  \bibfield  {author} {\bibinfo {author} {\bibfnamefont {J.~A.}\ \bibnamefont
  {Sauls}},\ }\bibfield  {title} {\bibinfo {title} {The order parameter for the
  superconducting phases of {UPt3}},\ }\href@noop {} {\bibfield  {journal}
  {\bibinfo  {journal} {Adv. Phys.}\ }\textbf {\bibinfo {volume} {43}},\
  \bibinfo {pages} {113} (\bibinfo {year} {1994})}\BibitemShut {NoStop}%
\bibitem [{\citenamefont {Joynt}\ and\ \citenamefont
  {Taillefer}(2002)}]{Joynt2002-xq}%
  \BibitemOpen
  \bibfield  {author} {\bibinfo {author} {\bibfnamefont {R.}~\bibnamefont
  {Joynt}}\ and\ \bibinfo {author} {\bibfnamefont {L.}~\bibnamefont
  {Taillefer}},\ }\bibfield  {title} {\bibinfo {title} {The superconducting
  phases of {UPt3}},\ }\href@noop {} {\bibfield  {journal} {\bibinfo  {journal}
  {Rev. Mod. Phys.}\ }\textbf {\bibinfo {volume} {74}},\ \bibinfo {pages} {235}
  (\bibinfo {year} {2002})}\BibitemShut {NoStop}%
\bibitem [{\citenamefont {How}\ and\ \citenamefont {Yip}(2019)}]{How2019-re}%
  \BibitemOpen
  \bibfield  {author} {\bibinfo {author} {\bibfnamefont {P.~T.}\ \bibnamefont
  {How}}\ and\ \bibinfo {author} {\bibfnamefont {S.-K.}\ \bibnamefont {Yip}},\
  }\bibfield  {title} {\bibinfo {title} {Signatures of nematic
  superconductivity in doped bi2se3 under applied stress},\ }\href@noop {}
  {\bibfield  {journal} {\bibinfo  {journal} {Phys. Rev. B: Condens. Matter
  Mater. Phys.}\ }\textbf {\bibinfo {volume} {100}},\ \bibinfo {pages} {134508}
  (\bibinfo {year} {2019})}\BibitemShut {NoStop}%
\bibitem [{Note1()}]{Note1}%
  \BibitemOpen
  \bibinfo {note} {Talk by Kristin Willa at Spin Phenomena Interdisciplinary
  Center, ``Evidence for nematic superconductivity in superconducting doped
  topological insulator Nb$_x$Bi$_2$Se$_3$ and Sr$_x$Bi$_2$Se$_3$'', available
  on https://www.youtube.com/watch?v=gVuzkKCU1xg}\BibitemShut {NoStop}%
\bibitem [{\citenamefont {Kostylev}\ \emph {et~al.}(2020)\citenamefont
  {Kostylev}, \citenamefont {Yonezawa}, \citenamefont {Wang}, \citenamefont
  {Ando},\ and\ \citenamefont {Maeno}}]{Kostylev2020-zb}%
  \BibitemOpen
  \bibfield  {author} {\bibinfo {author} {\bibfnamefont {I.}~\bibnamefont
  {Kostylev}}, \bibinfo {author} {\bibfnamefont {S.}~\bibnamefont {Yonezawa}},
  \bibinfo {author} {\bibfnamefont {Z.}~\bibnamefont {Wang}}, \bibinfo {author}
  {\bibfnamefont {Y.}~\bibnamefont {Ando}},\ and\ \bibinfo {author}
  {\bibfnamefont {Y.}~\bibnamefont {Maeno}},\ }\bibfield  {title} {\bibinfo
  {title} {Uniaxial-strain control of nematic superconductivity in
  {SrxBi2Se3}},\ }\href@noop {} {\bibfield  {journal} {\bibinfo  {journal}
  {Nat. Commun.}\ }\textbf {\bibinfo {volume} {11}},\ \bibinfo {pages} {4152}
  (\bibinfo {year} {2020})}\BibitemShut {NoStop}%
\bibitem [{\citenamefont {Willa}\ \emph {et~al.}(2018)\citenamefont {Willa},
  \citenamefont {Willa}, \citenamefont {Song}, \citenamefont {Gu},
  \citenamefont {Schneeloch}, \citenamefont {Zhong}, \citenamefont {Koshelev},
  \citenamefont {Kwok},\ and\ \citenamefont {Welp}}]{Willa2018-uu}%
  \BibitemOpen
  \bibfield  {author} {\bibinfo {author} {\bibfnamefont {K.}~\bibnamefont
  {Willa}}, \bibinfo {author} {\bibfnamefont {R.}~\bibnamefont {Willa}},
  \bibinfo {author} {\bibfnamefont {K.~W.}\ \bibnamefont {Song}}, \bibinfo
  {author} {\bibfnamefont {G.~D.}\ \bibnamefont {Gu}}, \bibinfo {author}
  {\bibfnamefont {J.~A.}\ \bibnamefont {Schneeloch}}, \bibinfo {author}
  {\bibfnamefont {R.}~\bibnamefont {Zhong}}, \bibinfo {author} {\bibfnamefont
  {A.~E.}\ \bibnamefont {Koshelev}}, \bibinfo {author} {\bibfnamefont {W.-K.}\
  \bibnamefont {Kwok}},\ and\ \bibinfo {author} {\bibfnamefont
  {U.}~\bibnamefont {Welp}},\ }\bibfield  {title} {\bibinfo {title}
  {Nanocalorimetric evidence for nematic superconductivity in the doped
  topological insulator {Sr0.1Bi2Se3}},\ }\href@noop {} {\bibfield  {journal}
  {\bibinfo  {journal} {Phys. Rev. B: Condens. Matter Mater. Phys.}\ }\textbf
  {\bibinfo {volume} {98}},\ \bibinfo {pages} {184509} (\bibinfo {year}
  {2018})}\BibitemShut {NoStop}%
\bibitem [{\citenamefont {Hess}\ \emph {et~al.}(1989)\citenamefont {Hess},
  \citenamefont {Tokuyasu},\ and\ \citenamefont {Sauls}}]{Hess1989-wi}%
  \BibitemOpen
  \bibfield  {author} {\bibinfo {author} {\bibfnamefont {D.~W.}\ \bibnamefont
  {Hess}}, \bibinfo {author} {\bibfnamefont {T.~A.}\ \bibnamefont {Tokuyasu}},\
  and\ \bibinfo {author} {\bibfnamefont {J.~A.}\ \bibnamefont {Sauls}},\
  }\bibfield  {title} {\bibinfo {title} {Broken symmetry in an unconventional
  superconductor: a model for the double transition in {UPt} 3},\ }\href@noop
  {} {\bibfield  {journal} {\bibinfo  {journal} {J. Phys. Condens. Matter}\
  }\textbf {\bibinfo {volume} {1}},\ \bibinfo {pages} {8135} (\bibinfo {year}
  {1989})}\BibitemShut {NoStop}%
\bibitem [{\citenamefont {Venderbos}\ \emph
  {et~al.}(2016{\natexlab{a}})\citenamefont {Venderbos}, \citenamefont
  {Kozii},\ and\ \citenamefont {Fu}}]{Venderbos2016-fa}%
  \BibitemOpen
  \bibfield  {author} {\bibinfo {author} {\bibfnamefont {J.~W.~F.}\
  \bibnamefont {Venderbos}}, \bibinfo {author} {\bibfnamefont {V.}~\bibnamefont
  {Kozii}},\ and\ \bibinfo {author} {\bibfnamefont {L.}~\bibnamefont {Fu}},\
  }\bibfield  {title} {\bibinfo {title} {Identification of nematic
  superconductivity from the upper critical field},\ }\href@noop {} {\bibfield
  {journal} {\bibinfo  {journal} {Phys. Rev. B: Condens. Matter Mater. Phys.}\
  }\textbf {\bibinfo {volume} {94}},\ \bibinfo {pages} {094522} (\bibinfo
  {year} {2016}{\natexlab{a}})}\BibitemShut {NoStop}%
\bibitem [{\citenamefont {Bannikov}\ \emph {et~al.}(2021)\citenamefont
  {Bannikov}, \citenamefont {Akzyanov}, \citenamefont {Zhurbina}, \citenamefont
  {Khaldeev}, \citenamefont {Selivanov}, \citenamefont {Zavyalov},
  \citenamefont {Rakhmanov},\ and\ \citenamefont
  {Kuntsevich}}]{Bannikov2021-uh}%
  \BibitemOpen
  \bibfield  {author} {\bibinfo {author} {\bibfnamefont {M.~I.}\ \bibnamefont
  {Bannikov}}, \bibinfo {author} {\bibfnamefont {R.~S.}\ \bibnamefont
  {Akzyanov}}, \bibinfo {author} {\bibfnamefont {N.~K.}\ \bibnamefont
  {Zhurbina}}, \bibinfo {author} {\bibfnamefont {S.~I.}\ \bibnamefont
  {Khaldeev}}, \bibinfo {author} {\bibfnamefont {Y.~G.}\ \bibnamefont
  {Selivanov}}, \bibinfo {author} {\bibfnamefont {V.~V.}\ \bibnamefont
  {Zavyalov}}, \bibinfo {author} {\bibfnamefont {A.~L.}\ \bibnamefont
  {Rakhmanov}},\ and\ \bibinfo {author} {\bibfnamefont {A.~Y.}\ \bibnamefont
  {Kuntsevich}},\ }\bibfield  {title} {\bibinfo {title} {Breaking of
  {Ginzburg-Landau} description in the temperature dependence of the anisotropy
  in a nematic superconductor},\ }\href@noop {} {\bibfield  {journal} {\bibinfo
   {journal} {Phys. Rev. B Condens. Matter}\ }\textbf {\bibinfo {volume}
  {104}},\ \bibinfo {pages} {L220502} (\bibinfo {year} {2021})}\BibitemShut
  {NoStop}%
\bibitem [{\citenamefont {Zyuzin}\ \emph {et~al.}(2017)\citenamefont {Zyuzin},
  \citenamefont {Garaud},\ and\ \citenamefont {Babaev}}]{Zyuzin2017-kn}%
  \BibitemOpen
  \bibfield  {author} {\bibinfo {author} {\bibfnamefont {A.~A.}\ \bibnamefont
  {Zyuzin}}, \bibinfo {author} {\bibfnamefont {J.}~\bibnamefont {Garaud}},\
  and\ \bibinfo {author} {\bibfnamefont {E.}~\bibnamefont {Babaev}},\
  }\bibfield  {title} {\bibinfo {title} {Nematic skyrmions in {Odd-Parity}
  superconductors},\ }\href@noop {} {\bibfield  {journal} {\bibinfo  {journal}
  {Phys. Rev. Lett.}\ }\textbf {\bibinfo {volume} {119}},\ \bibinfo {pages}
  {167001} (\bibinfo {year} {2017})}\BibitemShut {NoStop}%
\bibitem [{\citenamefont {How}\ and\ \citenamefont {Yip}(2020)}]{How2020-qj}%
  \BibitemOpen
  \bibfield  {author} {\bibinfo {author} {\bibfnamefont {P.~T.}\ \bibnamefont
  {How}}\ and\ \bibinfo {author} {\bibfnamefont {S.-K.}\ \bibnamefont {Yip}},\
  }\bibfield  {title} {\bibinfo {title} {Half quantum vortices in a nematic
  superconductor},\ }\href@noop {} {\bibfield  {journal} {\bibinfo  {journal}
  {Phys. Rev. Research}\ }\textbf {\bibinfo {volume} {2}},\ \bibinfo {pages}
  {043192} (\bibinfo {year} {2020})}\BibitemShut {NoStop}%
\bibitem [{\citenamefont {How}\ and\ \citenamefont {Yip}(2021)}]{How2021-me}%
  \BibitemOpen
  \bibfield  {author} {\bibinfo {author} {\bibfnamefont {P.~T.}\ \bibnamefont
  {How}}\ and\ \bibinfo {author} {\bibfnamefont {S.-K.}\ \bibnamefont {Yip}},\
  }\bibfield  {title} {\bibinfo {title} {Shear modulus anomaly of
  unconventional superconductors in a symmetry breaking field},\ }\href@noop {}
  {\bibfield  {journal} {\bibinfo  {journal} {Phys. Rev. B Condens. Matter}\
  }\textbf {\bibinfo {volume} {104}},\ \bibinfo {pages} {L020506} (\bibinfo
  {year} {2021})}\BibitemShut {NoStop}%
\bibitem [{\citenamefont {Hecker}\ and\ \citenamefont
  {Schmalian}(2018)}]{Hecker2018-bb}%
  \BibitemOpen
  \bibfield  {author} {\bibinfo {author} {\bibfnamefont {M.}~\bibnamefont
  {Hecker}}\ and\ \bibinfo {author} {\bibfnamefont {J.}~\bibnamefont
  {Schmalian}},\ }\bibfield  {title} {\bibinfo {title} {Vestigial nematic order
  and superconductivity in the doped topological insulator
  {Cu$_x$Bi$_2$Se$_3$}},\ }\href@noop {} {\bibfield  {journal} {\bibinfo
  {journal} {npj Quantum Materials}\ }\textbf {\bibinfo {volume} {3}},\
  \bibinfo {pages} {26} (\bibinfo {year} {2018})}\BibitemShut {NoStop}%
\bibitem [{\citenamefont {Sun}\ \emph {et~al.}(2019)\citenamefont {Sun},
  \citenamefont {Kittaka}, \citenamefont {Sakakibara}, \citenamefont {Machida},
  \citenamefont {Wang}, \citenamefont {Wen}, \citenamefont {Xing},
  \citenamefont {Shi},\ and\ \citenamefont {Tamegai}}]{Sun2019-kt}%
  \BibitemOpen
  \bibfield  {author} {\bibinfo {author} {\bibfnamefont {Y.}~\bibnamefont
  {Sun}}, \bibinfo {author} {\bibfnamefont {S.}~\bibnamefont {Kittaka}},
  \bibinfo {author} {\bibfnamefont {T.}~\bibnamefont {Sakakibara}}, \bibinfo
  {author} {\bibfnamefont {K.}~\bibnamefont {Machida}}, \bibinfo {author}
  {\bibfnamefont {J.}~\bibnamefont {Wang}}, \bibinfo {author} {\bibfnamefont
  {J.}~\bibnamefont {Wen}}, \bibinfo {author} {\bibfnamefont {X.}~\bibnamefont
  {Xing}}, \bibinfo {author} {\bibfnamefont {Z.}~\bibnamefont {Shi}},\ and\
  \bibinfo {author} {\bibfnamefont {T.}~\bibnamefont {Tamegai}},\ }\bibfield
  {title} {\bibinfo {title} {Quasiparticle evidence for the nematic state above
  tc in {Sr$_x$Bi$_2$Se$_3$}},\ }\href@noop {} {\bibfield  {journal} {\bibinfo
  {journal} {Phys. Rev. Lett.}\ }\textbf {\bibinfo {volume} {123}},\ \bibinfo
  {pages} {027002} (\bibinfo {year} {2019})}\BibitemShut {NoStop}%
\bibitem [{\citenamefont {Cho}\ \emph {et~al.}(2020)\citenamefont {Cho},
  \citenamefont {Shen}, \citenamefont {Lyu}, \citenamefont {Atanov},
  \citenamefont {Chen}, \citenamefont {Lee}, \citenamefont {Hor}, \citenamefont
  {Gawryluk}, \citenamefont {Pomjakushina}, \citenamefont {Bartkowiak},
  \citenamefont {Hecker}, \citenamefont {Schmalian},\ and\ \citenamefont
  {Lortz}}]{Cho2020-ne}%
  \BibitemOpen
  \bibfield  {author} {\bibinfo {author} {\bibfnamefont {C.-W.}\ \bibnamefont
  {Cho}}, \bibinfo {author} {\bibfnamefont {J.}~\bibnamefont {Shen}}, \bibinfo
  {author} {\bibfnamefont {J.}~\bibnamefont {Lyu}}, \bibinfo {author}
  {\bibfnamefont {O.}~\bibnamefont {Atanov}}, \bibinfo {author} {\bibfnamefont
  {Q.}~\bibnamefont {Chen}}, \bibinfo {author} {\bibfnamefont {S.~H.}\
  \bibnamefont {Lee}}, \bibinfo {author} {\bibfnamefont {Y.~S.}\ \bibnamefont
  {Hor}}, \bibinfo {author} {\bibfnamefont {D.~J.}\ \bibnamefont {Gawryluk}},
  \bibinfo {author} {\bibfnamefont {E.}~\bibnamefont {Pomjakushina}}, \bibinfo
  {author} {\bibfnamefont {M.}~\bibnamefont {Bartkowiak}}, \bibinfo {author}
  {\bibfnamefont {M.}~\bibnamefont {Hecker}}, \bibinfo {author} {\bibfnamefont
  {J.}~\bibnamefont {Schmalian}},\ and\ \bibinfo {author} {\bibfnamefont
  {R.}~\bibnamefont {Lortz}},\ }\bibfield  {title} {\bibinfo {title}
  {{$Z_3$}-vestigial nematic order due to superconducting fluctuations in the
  doped topological insulators {Nb$_x$Bi$_2$Se$_3$} and {Cu$_x$Bi$_2$Se$_3$}},\
  }\href@noop {} {\bibfield  {journal} {\bibinfo  {journal} {Nat. Commun.}\
  }\textbf {\bibinfo {volume} {11}},\ \bibinfo {pages} {3056} (\bibinfo {year}
  {2020})}\BibitemShut {NoStop}%
\bibitem [{\citenamefont {Kuntsevich}\ \emph {et~al.}(2018)\citenamefont
  {Kuntsevich}, \citenamefont {Bryzgalov}, \citenamefont {Prudkoglyad},
  \citenamefont {Martovitskii}, \citenamefont {Selivanov},\ and\ \citenamefont
  {Chizhevskii}}]{Kuntsevich2018-kk}%
  \BibitemOpen
  \bibfield  {author} {\bibinfo {author} {\bibfnamefont {A.~Y.}\ \bibnamefont
  {Kuntsevich}}, \bibinfo {author} {\bibfnamefont {M.~A.}\ \bibnamefont
  {Bryzgalov}}, \bibinfo {author} {\bibfnamefont {V.~A.}\ \bibnamefont
  {Prudkoglyad}}, \bibinfo {author} {\bibfnamefont {V.~P.}\ \bibnamefont
  {Martovitskii}}, \bibinfo {author} {\bibfnamefont {Y.~G.}\ \bibnamefont
  {Selivanov}},\ and\ \bibinfo {author} {\bibfnamefont {E.~G.}\ \bibnamefont
  {Chizhevskii}},\ }\bibfield  {title} {\bibinfo {title} {Structural distortion
  behind the nematic superconductivity in sr x bi 2 se 3},\ }\href@noop {}
  {\bibfield  {journal} {\bibinfo  {journal} {New J. Phys.}\ }\textbf {\bibinfo
  {volume} {20}},\ \bibinfo {pages} {103022} (\bibinfo {year}
  {2018})}\BibitemShut {NoStop}%
\bibitem [{\citenamefont {Bojesen}\ \emph {et~al.}(2014)\citenamefont
  {Bojesen}, \citenamefont {Babaev},\ and\ \citenamefont
  {Sudb{\o}}}]{Bojesen2014-vj}%
  \BibitemOpen
  \bibfield  {author} {\bibinfo {author} {\bibfnamefont {T.~A.}\ \bibnamefont
  {Bojesen}}, \bibinfo {author} {\bibfnamefont {E.}~\bibnamefont {Babaev}},\
  and\ \bibinfo {author} {\bibfnamefont {A.}~\bibnamefont {Sudb{\o}}},\
  }\bibfield  {title} {\bibinfo {title} {Phase transitions and anomalous normal
  state in superconductors with broken time-reversal symmetry},\ }\href@noop {}
  {\bibfield  {journal} {\bibinfo  {journal} {Phys. Rev. B Condens. Matter}\
  }\textbf {\bibinfo {volume} {89}},\ \bibinfo {pages} {104509} (\bibinfo
  {year} {2014})}\BibitemShut {NoStop}%
\bibitem [{\citenamefont {Fischer}\ and\ \citenamefont
  {Berg}(2016)}]{Fischer2016-gt}%
  \BibitemOpen
  \bibfield  {author} {\bibinfo {author} {\bibfnamefont {M.~H.}\ \bibnamefont
  {Fischer}}\ and\ \bibinfo {author} {\bibfnamefont {E.}~\bibnamefont {Berg}},\
  }\bibfield  {title} {\bibinfo {title} {Fluctuation and strain effects in a
  chiral $p$-wave superconductor},\ }\href@noop {} {\bibfield  {journal}
  {\bibinfo  {journal} {Phys. Rev. B Condens. Matter}\ }\textbf {\bibinfo
  {volume} {93}},\ \bibinfo {pages} {054501} (\bibinfo {year}
  {2016})}\BibitemShut {NoStop}%
\bibitem [{\citenamefont {Fernandes}\ \emph {et~al.}(2019)\citenamefont
  {Fernandes}, \citenamefont {Orth},\ and\ \citenamefont
  {Schmalian}}]{Fernandes2019-yv}%
  \BibitemOpen
  \bibfield  {author} {\bibinfo {author} {\bibfnamefont {R.~M.}\ \bibnamefont
  {Fernandes}}, \bibinfo {author} {\bibfnamefont {P.~P.}\ \bibnamefont
  {Orth}},\ and\ \bibinfo {author} {\bibfnamefont {J.}~\bibnamefont
  {Schmalian}},\ }\bibfield  {title} {\bibinfo {title} {Intertwined vestigial
  order in quantum materials: Nematicity and beyond},\ }\href@noop {}
  {\bibfield  {journal} {\bibinfo  {journal} {Annu. Rev. Condens. Matter
  Phys.}\ }\textbf {\bibinfo {volume} {10}},\ \bibinfo {pages} {133} (\bibinfo
  {year} {2019})}\BibitemShut {NoStop}%
\bibitem [{\citenamefont {Grinenko}\ \emph {et~al.}(2021)\citenamefont
  {Grinenko}, \citenamefont {Weston}, \citenamefont {Caglieris}, \citenamefont
  {Wuttke}, \citenamefont {Hess}, \citenamefont {Gottschall}, \citenamefont
  {Maccari}, \citenamefont {Gorbunov}, \citenamefont {Zherlitsyn},
  \citenamefont {Wosnitza}, \citenamefont {Rydh}, \citenamefont {Kihou},
  \citenamefont {Lee}, \citenamefont {Sarkar}, \citenamefont {Dengre},
  \citenamefont {Garaud}, \citenamefont {Charnukha}, \citenamefont {H{\"u}hne},
  \citenamefont {Nielsch}, \citenamefont {B{\"u}chner}, \citenamefont
  {Klauss},\ and\ \citenamefont {Babaev}}]{Grinenko2021-cw}%
  \BibitemOpen
  \bibfield  {author} {\bibinfo {author} {\bibfnamefont {V.}~\bibnamefont
  {Grinenko}}, \bibinfo {author} {\bibfnamefont {D.}~\bibnamefont {Weston}},
  \bibinfo {author} {\bibfnamefont {F.}~\bibnamefont {Caglieris}}, \bibinfo
  {author} {\bibfnamefont {C.}~\bibnamefont {Wuttke}}, \bibinfo {author}
  {\bibfnamefont {C.}~\bibnamefont {Hess}}, \bibinfo {author} {\bibfnamefont
  {T.}~\bibnamefont {Gottschall}}, \bibinfo {author} {\bibfnamefont
  {I.}~\bibnamefont {Maccari}}, \bibinfo {author} {\bibfnamefont
  {D.}~\bibnamefont {Gorbunov}}, \bibinfo {author} {\bibfnamefont
  {S.}~\bibnamefont {Zherlitsyn}}, \bibinfo {author} {\bibfnamefont
  {J.}~\bibnamefont {Wosnitza}}, \bibinfo {author} {\bibfnamefont
  {A.}~\bibnamefont {Rydh}}, \bibinfo {author} {\bibfnamefont {K.}~\bibnamefont
  {Kihou}}, \bibinfo {author} {\bibfnamefont {C.-H.}\ \bibnamefont {Lee}},
  \bibinfo {author} {\bibfnamefont {R.}~\bibnamefont {Sarkar}}, \bibinfo
  {author} {\bibfnamefont {S.}~\bibnamefont {Dengre}}, \bibinfo {author}
  {\bibfnamefont {J.}~\bibnamefont {Garaud}}, \bibinfo {author} {\bibfnamefont
  {A.}~\bibnamefont {Charnukha}}, \bibinfo {author} {\bibfnamefont
  {R.}~\bibnamefont {H{\"u}hne}}, \bibinfo {author} {\bibfnamefont
  {K.}~\bibnamefont {Nielsch}}, \bibinfo {author} {\bibfnamefont
  {B.}~\bibnamefont {B{\"u}chner}}, \bibinfo {author} {\bibfnamefont {H.-H.}\
  \bibnamefont {Klauss}},\ and\ \bibinfo {author} {\bibfnamefont
  {E.}~\bibnamefont {Babaev}},\ }\bibfield  {title} {\bibinfo {title} {State
  with spontaneously broken time-reversal symmetry above the superconducting
  phase transition},\ }\href@noop {} {\bibfield  {journal} {\bibinfo  {journal}
  {Nat. Phys.}\ }\textbf {\bibinfo {volume} {17}},\ \bibinfo {pages} {1254}
  (\bibinfo {year} {2021})}\BibitemShut {NoStop}%
\bibitem [{\citenamefont {Yuan}\ \emph {et~al.}(2017)\citenamefont {Yuan},
  \citenamefont {He},\ and\ \citenamefont {Law}}]{Yuan2017-la}%
  \BibitemOpen
  \bibfield  {author} {\bibinfo {author} {\bibfnamefont {N.~F.~Q.}\
  \bibnamefont {Yuan}}, \bibinfo {author} {\bibfnamefont {W.-Y.}\ \bibnamefont
  {He}},\ and\ \bibinfo {author} {\bibfnamefont {K.~T.}\ \bibnamefont {Law}},\
  }\bibfield  {title} {\bibinfo {title} {Superconductivity-induced
  ferromagnetism and weyl superconductivity in nb-doped {Bi$_{2}$Se$_{3}$}},\
  }\href@noop {} {\bibfield  {journal} {\bibinfo  {journal} {Phys. Rev. B
  Condens. Matter}\ }\textbf {\bibinfo {volume} {95}},\ \bibinfo {pages}
  {201109} (\bibinfo {year} {2017})}\BibitemShut {NoStop}%
\bibitem [{\citenamefont {Chirolli}(2018)}]{Chirolli2018-zj}%
  \BibitemOpen
  \bibfield  {author} {\bibinfo {author} {\bibfnamefont {L.}~\bibnamefont
  {Chirolli}},\ }\bibfield  {title} {\bibinfo {title} {Chiral superconductivity
  in thin films of doped {Bi$_{2}$Se$_{3}$}},\ }\href@noop {} {\bibfield
  {journal} {\bibinfo  {journal} {Phys. Rev. B Condens. Matter}\ }\textbf
  {\bibinfo {volume} {98}},\ \bibinfo {pages} {014505} (\bibinfo {year}
  {2018})}\BibitemShut {NoStop}%
\bibitem [{\citenamefont {Uematsu}\ \emph {et~al.}(2019)\citenamefont
  {Uematsu}, \citenamefont {Mizushima}, \citenamefont {Tsuruta}, \citenamefont
  {Fujimoto},\ and\ \citenamefont {Sauls}}]{Uematsu2019-ij}%
  \BibitemOpen
  \bibfield  {author} {\bibinfo {author} {\bibfnamefont {H.}~\bibnamefont
  {Uematsu}}, \bibinfo {author} {\bibfnamefont {T.}~\bibnamefont {Mizushima}},
  \bibinfo {author} {\bibfnamefont {A.}~\bibnamefont {Tsuruta}}, \bibinfo
  {author} {\bibfnamefont {S.}~\bibnamefont {Fujimoto}},\ and\ \bibinfo
  {author} {\bibfnamefont {J.~A.}\ \bibnamefont {Sauls}},\ }\bibfield  {title}
  {\bibinfo {title} {Chiral higgs mode in nematic superconductors},\
  }\href@noop {} {\bibfield  {journal} {\bibinfo  {journal} {Phys. Rev. Lett.}\
  }\textbf {\bibinfo {volume} {123}},\ \bibinfo {pages} {237001} (\bibinfo
  {year} {2019})}\BibitemShut {NoStop}%
\bibitem [{\citenamefont {How}\ and\ \citenamefont {Yip}(2022)}]{How2022-qb}%
  \BibitemOpen
  \bibfield  {author} {\bibinfo {author} {\bibfnamefont {P.~T.}\ \bibnamefont
  {How}}\ and\ \bibinfo {author} {\bibfnamefont {S.~K.}\ \bibnamefont {Yip}},\
  }\bibfield  {title} {\bibinfo {title} {Criterion for vestigial order above a
  nematic superconductor},\ }\href@noop {} {\  (\bibinfo {year} {2022})},\
  \Eprint {https://arxiv.org/abs/2207.04714} {arXiv:2207.04714
  [cond-mat.supr-con]} \BibitemShut {NoStop}%
\bibitem [{\citenamefont {Barash}\ and\ \citenamefont
  {Galaktionov}(1991)}]{Barash1991-qg}%
  \BibitemOpen
  \bibfield  {author} {\bibinfo {author} {\bibfnamefont {Y.~S.}\ \bibnamefont
  {Barash}}\ and\ \bibinfo {author} {\bibfnamefont {A.~V.}\ \bibnamefont
  {Galaktionov}},\ }\bibfield  {title} {\bibinfo {title} {Anisotropy of
  magnetic properties of exotic superconductors},\ }\href@noop {} {\bibfield
  {journal} {\bibinfo  {journal} {Zh. Eksp. Teor. Fiz.}\ }\textbf {\bibinfo
  {volume} {100}},\ \bibinfo {pages} {1699} (\bibinfo {year}
  {1991})}\BibitemShut {NoStop}%
\bibitem [{\citenamefont {Akzyanov}\ \emph {et~al.}(2020)\citenamefont
  {Akzyanov}, \citenamefont {Khokhlov},\ and\ \citenamefont
  {Rakhmanov}}]{Akzyanov2020-zb}%
  \BibitemOpen
  \bibfield  {author} {\bibinfo {author} {\bibfnamefont {R.~S.}\ \bibnamefont
  {Akzyanov}}, \bibinfo {author} {\bibfnamefont {D.~A.}\ \bibnamefont
  {Khokhlov}},\ and\ \bibinfo {author} {\bibfnamefont {A.~L.}\ \bibnamefont
  {Rakhmanov}},\ }\bibfield  {title} {\bibinfo {title} {Nematic
  superconductivity in topological insulators induced by hexagonal warping},\
  }\href@noop {} {\bibfield  {journal} {\bibinfo  {journal} {Phys. Rev. B
  Condens. Matter}\ }\textbf {\bibinfo {volume} {102}},\ \bibinfo {pages}
  {094511} (\bibinfo {year} {2020})}\BibitemShut {NoStop}%
\bibitem [{Note2()}]{Note2}%
  \BibitemOpen
  \bibinfo {note} {If one attempts to construct this effective theory starting
  from the approximate $k \cdot p$ electron Hamiltonian of $M_x$Bi$_2$Se$_3$
  \cite {Zyuzin2017-kn}, then $K'$ (accidentally) vanishes. This is an artifact
  of the $k \cdot p$ approximation.}\BibitemShut {Stop}%
\bibitem [{\citenamefont {Babaev}\ \emph {et~al.}(2004)\citenamefont {Babaev},
  \citenamefont {Sudb{\o}},\ and\ \citenamefont {Ashcroft}}]{Babaev2004-op}%
  \BibitemOpen
  \bibfield  {author} {\bibinfo {author} {\bibfnamefont {E.}~\bibnamefont
  {Babaev}}, \bibinfo {author} {\bibfnamefont {A.}~\bibnamefont {Sudb{\o}}},\
  and\ \bibinfo {author} {\bibfnamefont {N.~W.}\ \bibnamefont {Ashcroft}},\
  }\bibfield  {title} {\bibinfo {title} {A superconductor to superfluid phase
  transition in liquid metallic hydrogen},\ }\href@noop {} {\bibfield
  {journal} {\bibinfo  {journal} {Nature}\ }\textbf {\bibinfo {volume} {431}},\
  \bibinfo {pages} {666} (\bibinfo {year} {2004})}\BibitemShut {NoStop}%
\bibitem [{\citenamefont {Kuklov}\ \emph {et~al.}(2004)\citenamefont {Kuklov},
  \citenamefont {Prokof'ev},\ and\ \citenamefont {Svistunov}}]{Kuklov2004-ib}%
  \BibitemOpen
  \bibfield  {author} {\bibinfo {author} {\bibfnamefont {A.}~\bibnamefont
  {Kuklov}}, \bibinfo {author} {\bibfnamefont {N.}~\bibnamefont {Prokof'ev}},\
  and\ \bibinfo {author} {\bibfnamefont {B.}~\bibnamefont {Svistunov}},\
  }\bibfield  {title} {\bibinfo {title} {Commensurate {Two-Component} bosons in
  an optical lattice: Ground state phase diagram},\ }\href@noop {} {\bibfield
  {journal} {\bibinfo  {journal} {Phys. Rev. Lett.}\ }\textbf {\bibinfo
  {volume} {92}},\ \bibinfo {pages} {050402} (\bibinfo {year}
  {2004})}\BibitemShut {NoStop}%
\bibitem [{\citenamefont {Kuklov}\ \emph {et~al.}(2008)\citenamefont {Kuklov},
  \citenamefont {Matsumoto}, \citenamefont {Prokof'ev}, \citenamefont
  {Svistunov},\ and\ \citenamefont {Troyer}}]{Kuklov2008-oz}%
  \BibitemOpen
  \bibfield  {author} {\bibinfo {author} {\bibfnamefont {A.~B.}\ \bibnamefont
  {Kuklov}}, \bibinfo {author} {\bibfnamefont {M.}~\bibnamefont {Matsumoto}},
  \bibinfo {author} {\bibfnamefont {N.~V.}\ \bibnamefont {Prokof'ev}}, \bibinfo
  {author} {\bibfnamefont {B.~V.}\ \bibnamefont {Svistunov}},\ and\ \bibinfo
  {author} {\bibfnamefont {M.}~\bibnamefont {Troyer}},\ }\bibfield  {title}
  {\bibinfo {title} {Deconfined criticality: Generic {First-Order} transition
  in the {SU(2}) symmetry case},\ }\href@noop {} {\bibfield  {journal}
  {\bibinfo  {journal} {Phys. Rev. Lett.}\ }\textbf {\bibinfo {volume} {101}},\
  \bibinfo {pages} {050405} (\bibinfo {year} {2008})}\BibitemShut {NoStop}%
\bibitem [{\citenamefont {Herland}\ \emph {et~al.}(2010)\citenamefont
  {Herland}, \citenamefont {Babaev},\ and\ \citenamefont
  {Sudb{\o}}}]{Herland2010-di}%
  \BibitemOpen
  \bibfield  {author} {\bibinfo {author} {\bibfnamefont {E.~V.}\ \bibnamefont
  {Herland}}, \bibinfo {author} {\bibfnamefont {E.}~\bibnamefont {Babaev}},\
  and\ \bibinfo {author} {\bibfnamefont {A.}~\bibnamefont {Sudb{\o}}},\
  }\bibfield  {title} {\bibinfo {title} {Phase transitions in a three
  dimensional {$U(1)\times U(1)$} lattice london superconductor: Metallic
  superfluid and charge-$4e$ superconducting states},\ }\href@noop {}
  {\bibfield  {journal} {\bibinfo  {journal} {Phys. Rev. B Condens. Matter}\
  }\textbf {\bibinfo {volume} {82}},\ \bibinfo {pages} {134511} (\bibinfo
  {year} {2010})}\BibitemShut {NoStop}%
\bibitem [{\citenamefont {Bojesen}\ \emph {et~al.}(2013)\citenamefont
  {Bojesen}, \citenamefont {Babaev},\ and\ \citenamefont
  {Sudb{\o}}}]{Bojesen2013-lt}%
  \BibitemOpen
  \bibfield  {author} {\bibinfo {author} {\bibfnamefont {T.~A.}\ \bibnamefont
  {Bojesen}}, \bibinfo {author} {\bibfnamefont {E.}~\bibnamefont {Babaev}},\
  and\ \bibinfo {author} {\bibfnamefont {A.}~\bibnamefont {Sudb{\o}}},\
  }\bibfield  {title} {\bibinfo {title} {Time reversal symmetry breakdown in
  normal and superconducting states in frustrated three-band systems},\
  }\href@noop {} {\bibfield  {journal} {\bibinfo  {journal} {Phys. Rev. B
  Condens. Matter}\ }\textbf {\bibinfo {volume} {88}},\ \bibinfo {pages}
  {220511} (\bibinfo {year} {2013})}\BibitemShut {NoStop}%
\bibitem [{Note3()}]{Note3}%
  \BibitemOpen
  \bibinfo {note} {Subsequent discussions are often framed in terms of the
  ordering in \protect \emph {relative phase} between field components, and
  thus bear less superficial resemblance to the present scenario. And of course
  the models and the mechanisms discussed there are very
  different.}\BibitemShut {Stop}%
\bibitem [{\citenamefont {Venderbos}\ \emph
  {et~al.}(2016{\natexlab{b}})\citenamefont {Venderbos}, \citenamefont
  {Kozii},\ and\ \citenamefont {Fu}}]{Venderbos2016-cv}%
  \BibitemOpen
  \bibfield  {author} {\bibinfo {author} {\bibfnamefont {J.~W.~F.}\
  \bibnamefont {Venderbos}}, \bibinfo {author} {\bibfnamefont {V.}~\bibnamefont
  {Kozii}},\ and\ \bibinfo {author} {\bibfnamefont {L.}~\bibnamefont {Fu}},\
  }\bibfield  {title} {\bibinfo {title} {Odd-parity superconductors with
  two-component order parameters: Nematic and chiral, full gap, and majorana
  node},\ }\href@noop {} {\bibfield  {journal} {\bibinfo  {journal} {Phys. Rev.
  B: Condens. Matter Mater. Phys.}\ }\textbf {\bibinfo {volume} {94}},\
  \bibinfo {pages} {180504} (\bibinfo {year} {2016}{\natexlab{b}})}\BibitemShut
  {NoStop}%
\bibitem [{Note4()}]{Note4}%
  \BibitemOpen
  \bibinfo {note} {Ref \cite {Venderbos2016-cv} wrote the quartic terms of the
  free energy in a different form: their $B_1$ and $B_2$ are proportional to
  our $\beta _1 + \beta _2$ and $-\beta _2$, respectively. The sign of our
  $g_2$ is thus the same as that of $B_1 - B_2$ in their supplementary
  material. It seems to us that this quantity is always non-negative for every
  model discussed there.}\BibitemShut {Stop}%
\bibitem [{\citenamefont {Luttinger}\ and\ \citenamefont
  {Ward}(1960)}]{Luttinger1960-el}%
  \BibitemOpen
  \bibfield  {author} {\bibinfo {author} {\bibfnamefont {J.~M.}\ \bibnamefont
  {Luttinger}}\ and\ \bibinfo {author} {\bibfnamefont {J.~C.}\ \bibnamefont
  {Ward}},\ }\bibfield  {title} {\bibinfo {title} {{Ground-State} energy of a
  {Many-Fermion} system. {II}},\ }\href@noop {} {\bibfield  {journal} {\bibinfo
   {journal} {Phys. Rev.}\ }\textbf {\bibinfo {volume} {118}},\ \bibinfo
  {pages} {1417} (\bibinfo {year} {1960})}\BibitemShut {NoStop}%
\bibitem [{\citenamefont {Baym}\ and\ \citenamefont
  {Grinstein}(1977)}]{Baym1977-lm}%
  \BibitemOpen
  \bibfield  {author} {\bibinfo {author} {\bibfnamefont {G.}~\bibnamefont
  {Baym}}\ and\ \bibinfo {author} {\bibfnamefont {G.}~\bibnamefont
  {Grinstein}},\ }\bibfield  {title} {\bibinfo {title} {Phase transition in the
  sigma model at finite temperature},\ }\href@noop {} {\bibfield  {journal}
  {\bibinfo  {journal} {Physical Review D}\ }\textbf {\bibinfo {volume} {15}},\
  \bibinfo {pages} {2897} (\bibinfo {year} {1977})}\BibitemShut {NoStop}%
\bibitem [{\citenamefont {van Hees}\ and\ \citenamefont
  {Knoll}(2001)}]{Van_Hees2001-hl}%
  \BibitemOpen
  \bibfield  {author} {\bibinfo {author} {\bibfnamefont {H.}~\bibnamefont {van
  Hees}}\ and\ \bibinfo {author} {\bibfnamefont {J.}~\bibnamefont {Knoll}},\
  }\bibfield  {title} {\bibinfo {title} {Renormalization in self-consistent
  approximation schemes at finite temperature: Theory},\ }\href@noop {}
  {\bibfield  {journal} {\bibinfo  {journal} {Phys. Rev. D}\ }\textbf {\bibinfo
  {volume} {65}},\ \bibinfo {pages} {025010} (\bibinfo {year}
  {2001})}\BibitemShut {NoStop}%
\bibitem [{\citenamefont {Blaizot}\ \emph {et~al.}(2004)\citenamefont
  {Blaizot}, \citenamefont {Iancu},\ and\ \citenamefont
  {Reinosa}}]{Blaizot2004-ae}%
  \BibitemOpen
  \bibfield  {author} {\bibinfo {author} {\bibfnamefont {J.-P.}\ \bibnamefont
  {Blaizot}}, \bibinfo {author} {\bibfnamefont {E.}~\bibnamefont {Iancu}},\
  and\ \bibinfo {author} {\bibfnamefont {U.}~\bibnamefont {Reinosa}},\
  }\bibfield  {title} {\bibinfo {title} {Renormalization of {$\Phi$-derivable}
  approximations in scalar field theories},\ }\href@noop {} {\bibfield
  {journal} {\bibinfo  {journal} {Nucl. Phys. A}\ }\textbf {\bibinfo {volume}
  {736}},\ \bibinfo {pages} {149} (\bibinfo {year} {2004})}\BibitemShut
  {NoStop}%
\bibitem [{\citenamefont {Berges}\ \emph {et~al.}(2005)\citenamefont {Berges},
  \citenamefont {Bors{\'a}nyi}, \citenamefont {Reinosa},\ and\ \citenamefont
  {Serreau}}]{Berges2005-yz}%
  \BibitemOpen
  \bibfield  {author} {\bibinfo {author} {\bibfnamefont {J.}~\bibnamefont
  {Berges}}, \bibinfo {author} {\bibfnamefont {S.}~\bibnamefont
  {Bors{\'a}nyi}}, \bibinfo {author} {\bibfnamefont {U.}~\bibnamefont
  {Reinosa}},\ and\ \bibinfo {author} {\bibfnamefont {J.}~\bibnamefont
  {Serreau}},\ }\bibfield  {title} {\bibinfo {title} {Nonperturbative
  renormalization for {2PI} effective action techniques},\ }\href@noop {}
  {\bibfield  {journal} {\bibinfo  {journal} {Ann. Phys.}\ }\textbf {\bibinfo
  {volume} {320}},\ \bibinfo {pages} {344} (\bibinfo {year}
  {2005})}\BibitemShut {NoStop}%
\bibitem [{\citenamefont {Fradkin}(2021)}]{Fradkin2021-zu}%
  \BibitemOpen
  \bibfield  {author} {\bibinfo {author} {\bibfnamefont {E.}~\bibnamefont
  {Fradkin}},\ }\href@noop {} {\emph {\bibinfo {title} {Quantum field theory:
  An integrated approach}}}\ (\bibinfo  {publisher} {Princeton University
  Press},\ \bibinfo {address} {Princeton, NJ},\ \bibinfo {year}
  {2021})\BibitemShut {NoStop}%
\bibitem [{\citenamefont {H{\"u}gel}\ \emph {et~al.}(2016)\citenamefont
  {H{\"u}gel}, \citenamefont {Werner}, \citenamefont {Pollet},\ and\
  \citenamefont {Strand}}]{Hugel2016-bx}%
  \BibitemOpen
  \bibfield  {author} {\bibinfo {author} {\bibfnamefont {D.}~\bibnamefont
  {H{\"u}gel}}, \bibinfo {author} {\bibfnamefont {P.}~\bibnamefont {Werner}},
  \bibinfo {author} {\bibfnamefont {L.}~\bibnamefont {Pollet}},\ and\ \bibinfo
  {author} {\bibfnamefont {H.~U.~R.}\ \bibnamefont {Strand}},\ }\bibfield
  {title} {\bibinfo {title} {Bosonic self-energy functional theory},\
  }\href@noop {} {\bibfield  {journal} {\bibinfo  {journal} {Phys. Rev. B
  Condens. Matter}\ }\textbf {\bibinfo {volume} {94}},\ \bibinfo {pages}
  {195119} (\bibinfo {year} {2016})}\BibitemShut {NoStop}%
\bibitem [{Note5()}]{Note5}%
  \BibitemOpen
  \bibinfo {note} {The omission of the $h_0$ variation in our previous preprint
  \cite {How2022-qb} was the single mistake that led to its erroneous
  conclusion. It amounted to an extra constraint placed on the system. With
  $h_0$ artificially set to zero here, one can recover the result of \cite
  {How2022-qb}.}\BibitemShut {Stop}%
\bibitem [{Note6()}]{Note6}%
  \BibitemOpen
  \bibinfo {note} {Closed form expression of $\alpha _{c2}$ can be obtained by
  solving the saddle point equations with the additional condition $u = v$. It
  is unwieldy and not particularly useful, and we choose to omit the explicit
  expression here.}\BibitemShut {Stop}%
\bibitem [{\citenamefont {Patashinskii}\ and\ \citenamefont
  {Pokrovskii}(1979)}]{Patashinskii1979-hr}%
  \BibitemOpen
  \bibfield  {author} {\bibinfo {author} {\bibfnamefont {A.~Z.}\ \bibnamefont
  {Patashinskii}}\ and\ \bibinfo {author} {\bibfnamefont {V.~L.}\ \bibnamefont
  {Pokrovskii}},\ }\href@noop {} {\emph {\bibinfo {title} {Fluctuation Theory
  of Phase Transitions}}},\ Monographs in Natural Philosophy\ (\bibinfo
  {publisher} {Pergamon Press},\ \bibinfo {year} {1979})\BibitemShut {NoStop}%
\bibitem [{\citenamefont {Kawamura}(1986)}]{Kawamura1986-mo}%
  \BibitemOpen
  \bibfield  {author} {\bibinfo {author} {\bibfnamefont {H.}~\bibnamefont
  {Kawamura}},\ }\bibfield  {title} {\bibinfo {title} {{Renormalization-Group}
  approach to the frustrated heisenberg antiferromagnet on the
  {Layered-Triangular} lattice},\ }\href@noop {} {\bibfield  {journal}
  {\bibinfo  {journal} {J. Phys. Soc. Jpn.}\ }\textbf {\bibinfo {volume}
  {55}},\ \bibinfo {pages} {2157} (\bibinfo {year} {1986})}\BibitemShut
  {NoStop}%
\bibitem [{\citenamefont {Kawamura}(1988)}]{Kawamura1988-ga}%
  \BibitemOpen
  \bibfield  {author} {\bibinfo {author} {\bibfnamefont {H.}~\bibnamefont
  {Kawamura}},\ }\bibfield  {title} {\bibinfo {title} {Renormalization-group
  analysis of chiral transitions},\ }\href@noop {} {\bibfield  {journal}
  {\bibinfo  {journal} {Phys. Rev. B Condens. Matter}\ }\textbf {\bibinfo
  {volume} {38}},\ \bibinfo {pages} {4916} (\bibinfo {year}
  {1988})}\BibitemShut {NoStop}%
\bibitem [{\citenamefont {Antonenko}\ \emph {et~al.}(1995)\citenamefont
  {Antonenko}, \citenamefont {Sokolov},\ and\ \citenamefont
  {Varnashev}}]{Antonenko1995-ix}%
  \BibitemOpen
  \bibfield  {author} {\bibinfo {author} {\bibfnamefont {S.~A.}\ \bibnamefont
  {Antonenko}}, \bibinfo {author} {\bibfnamefont {A.~I.}\ \bibnamefont
  {Sokolov}},\ and\ \bibinfo {author} {\bibfnamefont {K.~B.}\ \bibnamefont
  {Varnashev}},\ }\bibfield  {title} {\bibinfo {title} {Chiral transitions in
  three-dimensional magnets and higher order $\epsilon$-expansion},\
  }\href@noop {} {\bibfield  {journal} {\bibinfo  {journal} {Phys. Lett. A}\
  }\textbf {\bibinfo {volume} {208}},\ \bibinfo {pages} {161} (\bibinfo {year}
  {1995})}\BibitemShut {NoStop}%
\bibitem [{\citenamefont {Delamotte}\ \emph {et~al.}(2004)\citenamefont
  {Delamotte}, \citenamefont {Mouhanna},\ and\ \citenamefont
  {Tissier}}]{Delamotte2004-rv}%
  \BibitemOpen
  \bibfield  {author} {\bibinfo {author} {\bibfnamefont {B.}~\bibnamefont
  {Delamotte}}, \bibinfo {author} {\bibfnamefont {D.}~\bibnamefont
  {Mouhanna}},\ and\ \bibinfo {author} {\bibfnamefont {M.}~\bibnamefont
  {Tissier}},\ }\bibfield  {title} {\bibinfo {title} {Nonperturbative
  renormalization-group approach to frustrated magnets},\ }\href@noop {}
  {\bibfield  {journal} {\bibinfo  {journal} {Phys. Rev. B Condens. Matter}\
  }\textbf {\bibinfo {volume} {69}},\ \bibinfo {pages} {134413} (\bibinfo
  {year} {2004})}\BibitemShut {NoStop}%
\bibitem [{\citenamefont {Sorokin}(2019)}]{Sorokin2019-ed}%
  \BibitemOpen
  \bibfield  {author} {\bibinfo {author} {\bibfnamefont {A.~O.}\ \bibnamefont
  {Sorokin}},\ }\bibfield  {title} {\bibinfo {title} {Weak {First-Order}
  transition and pseudoscaling behavior in the universality class of the {O(N})
  ising model},\ }\href@noop {} {\bibfield  {journal} {\bibinfo  {journal}
  {Theor. Math. Phys.}\ }\textbf {\bibinfo {volume} {200}},\ \bibinfo {pages}
  {1193} (\bibinfo {year} {2019})}\BibitemShut {NoStop}%
\bibitem [{\citenamefont {Eder}(2019)}]{Eder2019-il}%
  \BibitemOpen
  \bibfield  {author} {\bibinfo {author} {\bibfnamefont {R.}~\bibnamefont
  {Eder}},\ }\bibfield  {title} {\bibinfo {title} {Analytical properties of
  {Self-Energy} and {Luttinger-Ward} functional},\ }in\ \href@noop {} {\emph
  {\bibinfo {booktitle} {{Many-Body} Methods for Real Materials (Lecture notes
  of the Autumn School on Correlated Electrons 2019)}}},\ \bibinfo {series}
  {Modeling and Simulation}, Vol.~\bibinfo {volume} {9},\ \bibinfo {editor}
  {edited by\ \bibinfo {editor} {\bibfnamefont {E.}~\bibnamefont {Pavarini}},
  \bibinfo {editor} {\bibfnamefont {E.}~\bibnamefont {Koch}},\ and\ \bibinfo
  {editor} {\bibfnamefont {S.}~\bibnamefont {Zhang}}}\ (\bibinfo {year}
  {2019})\BibitemShut {NoStop}%
\bibitem [{\citenamefont {Lahoud}\ \emph {et~al.}(2013)\citenamefont {Lahoud},
  \citenamefont {Maniv}, \citenamefont {Petrushevsky}, \citenamefont {Naamneh},
  \citenamefont {Ribak}, \citenamefont {Wiedmann}, \citenamefont {Petaccia},
  \citenamefont {Salman}, \citenamefont {Chashka}, \citenamefont {Dagan},\ and\
  \citenamefont {Kanigel}}]{Lahoud2013-rl}%
  \BibitemOpen
  \bibfield  {author} {\bibinfo {author} {\bibfnamefont {E.}~\bibnamefont
  {Lahoud}}, \bibinfo {author} {\bibfnamefont {E.}~\bibnamefont {Maniv}},
  \bibinfo {author} {\bibfnamefont {M.~S.}\ \bibnamefont {Petrushevsky}},
  \bibinfo {author} {\bibfnamefont {M.}~\bibnamefont {Naamneh}}, \bibinfo
  {author} {\bibfnamefont {A.}~\bibnamefont {Ribak}}, \bibinfo {author}
  {\bibfnamefont {S.}~\bibnamefont {Wiedmann}}, \bibinfo {author}
  {\bibfnamefont {L.}~\bibnamefont {Petaccia}}, \bibinfo {author}
  {\bibfnamefont {Z.}~\bibnamefont {Salman}}, \bibinfo {author} {\bibfnamefont
  {K.~B.}\ \bibnamefont {Chashka}}, \bibinfo {author} {\bibfnamefont
  {Y.}~\bibnamefont {Dagan}},\ and\ \bibinfo {author} {\bibfnamefont
  {A.}~\bibnamefont {Kanigel}},\ }\bibfield  {title} {\bibinfo {title}
  {Evolution of the fermi surface of a doped topological insulator with carrier
  concentration},\ }\href@noop {} {\bibfield  {journal} {\bibinfo  {journal}
  {Phys. Rev. B: Condens. Matter Mater. Phys.}\ }\textbf {\bibinfo {volume}
  {88}},\ \bibinfo {pages} {195107} (\bibinfo {year} {2013})}\BibitemShut
  {NoStop}%
\bibitem [{\citenamefont {Almoalem}\ \emph {et~al.}(2021)\citenamefont
  {Almoalem}, \citenamefont {Silber}, \citenamefont {Sandik}, \citenamefont
  {Lotem}, \citenamefont {Ribak}, \citenamefont {Nitzav}, \citenamefont
  {Kuntsevich}, \citenamefont {Sobolevskiy}, \citenamefont {Selivanov},
  \citenamefont {Prudkoglyad}, \citenamefont {Shi}, \citenamefont {Petaccia},
  \citenamefont {Goldstein}, \citenamefont {Dagan},\ and\ \citenamefont
  {Kanigel}}]{Almoalem2021-zf}%
  \BibitemOpen
  \bibfield  {author} {\bibinfo {author} {\bibfnamefont {A.}~\bibnamefont
  {Almoalem}}, \bibinfo {author} {\bibfnamefont {I.}~\bibnamefont {Silber}},
  \bibinfo {author} {\bibfnamefont {S.}~\bibnamefont {Sandik}}, \bibinfo
  {author} {\bibfnamefont {M.}~\bibnamefont {Lotem}}, \bibinfo {author}
  {\bibfnamefont {A.}~\bibnamefont {Ribak}}, \bibinfo {author} {\bibfnamefont
  {Y.}~\bibnamefont {Nitzav}}, \bibinfo {author} {\bibfnamefont {A.~Y.}\
  \bibnamefont {Kuntsevich}}, \bibinfo {author} {\bibfnamefont {O.~A.}\
  \bibnamefont {Sobolevskiy}}, \bibinfo {author} {\bibfnamefont {Y.~G.}\
  \bibnamefont {Selivanov}}, \bibinfo {author} {\bibfnamefont {V.~A.}\
  \bibnamefont {Prudkoglyad}}, \bibinfo {author} {\bibfnamefont
  {M.}~\bibnamefont {Shi}}, \bibinfo {author} {\bibfnamefont {L.}~\bibnamefont
  {Petaccia}}, \bibinfo {author} {\bibfnamefont {M.}~\bibnamefont {Goldstein}},
  \bibinfo {author} {\bibfnamefont {Y.}~\bibnamefont {Dagan}},\ and\ \bibinfo
  {author} {\bibfnamefont {A.}~\bibnamefont {Kanigel}},\ }\bibfield  {title}
  {\bibinfo {title} {Link between superconductivity and a lifshitz transition
  in intercalated {Bi$_{2}$Se$_{3}$}},\ }\href@noop {} {\bibfield  {journal}
  {\bibinfo  {journal} {Phys. Rev. B Condens. Matter}\ }\textbf {\bibinfo
  {volume} {103}},\ \bibinfo {pages} {174518} (\bibinfo {year}
  {2021})}\BibitemShut {NoStop}%
\bibitem [{\citenamefont {Kriener}\ \emph {et~al.}(2011)\citenamefont
  {Kriener}, \citenamefont {Segawa}, \citenamefont {Ren}, \citenamefont
  {Sasaki},\ and\ \citenamefont {Ando}}]{Kriener2011-ha}%
  \BibitemOpen
  \bibfield  {author} {\bibinfo {author} {\bibfnamefont {M.}~\bibnamefont
  {Kriener}}, \bibinfo {author} {\bibfnamefont {K.}~\bibnamefont {Segawa}},
  \bibinfo {author} {\bibfnamefont {Z.}~\bibnamefont {Ren}}, \bibinfo {author}
  {\bibfnamefont {S.}~\bibnamefont {Sasaki}},\ and\ \bibinfo {author}
  {\bibfnamefont {Y.}~\bibnamefont {Ando}},\ }\bibfield  {title} {\bibinfo
  {title} {Bulk superconducting phase with a full energy gap in the doped
  topological insulator {CuxBi2Se3}},\ }\href@noop {} {\bibfield  {journal}
  {\bibinfo  {journal} {Phys. Rev. Lett.}\ }\textbf {\bibinfo {volume} {106}},\
  \bibinfo {pages} {127004} (\bibinfo {year} {2011})}\BibitemShut {NoStop}%
\bibitem [{\citenamefont {Wray}\ \emph {et~al.}(2010)\citenamefont {Wray},
  \citenamefont {Xu}, \citenamefont {Xia}, \citenamefont {Hor}, \citenamefont
  {Qian}, \citenamefont {Fedorov}, \citenamefont {Lin}, \citenamefont {Bansil},
  \citenamefont {Cava},\ and\ \citenamefont {Hasan}}]{Wray2010-ct}%
  \BibitemOpen
  \bibfield  {author} {\bibinfo {author} {\bibfnamefont {L.~A.}\ \bibnamefont
  {Wray}}, \bibinfo {author} {\bibfnamefont {S.-Y.}\ \bibnamefont {Xu}},
  \bibinfo {author} {\bibfnamefont {Y.}~\bibnamefont {Xia}}, \bibinfo {author}
  {\bibfnamefont {Y.~S.}\ \bibnamefont {Hor}}, \bibinfo {author} {\bibfnamefont
  {D.}~\bibnamefont {Qian}}, \bibinfo {author} {\bibfnamefont {A.~V.}\
  \bibnamefont {Fedorov}}, \bibinfo {author} {\bibfnamefont {H.}~\bibnamefont
  {Lin}}, \bibinfo {author} {\bibfnamefont {A.}~\bibnamefont {Bansil}},
  \bibinfo {author} {\bibfnamefont {R.~J.}\ \bibnamefont {Cava}},\ and\
  \bibinfo {author} {\bibfnamefont {M.~Z.}\ \bibnamefont {Hasan}},\ }\bibfield
  {title} {\bibinfo {title} {Observation of topological order in a
  superconducting doped topological insulator},\ }\href@noop {} {\bibfield
  {journal} {\bibinfo  {journal} {Nat. Phys.}\ }\textbf {\bibinfo {volume}
  {6}},\ \bibinfo {pages} {855} (\bibinfo {year} {2010})}\BibitemShut {NoStop}%
\end{thebibliography}%

\appendix

\section{Renormalized Bosonic Luttinger-Ward Formalism}

In this appendix, we justify the treatment of the counter term $\delta \alpha$, and argue that the result is UV-finite to all order of approximation.

\begin{figure}
\subfloat[]{\includegraphics[width=0.2\textwidth]{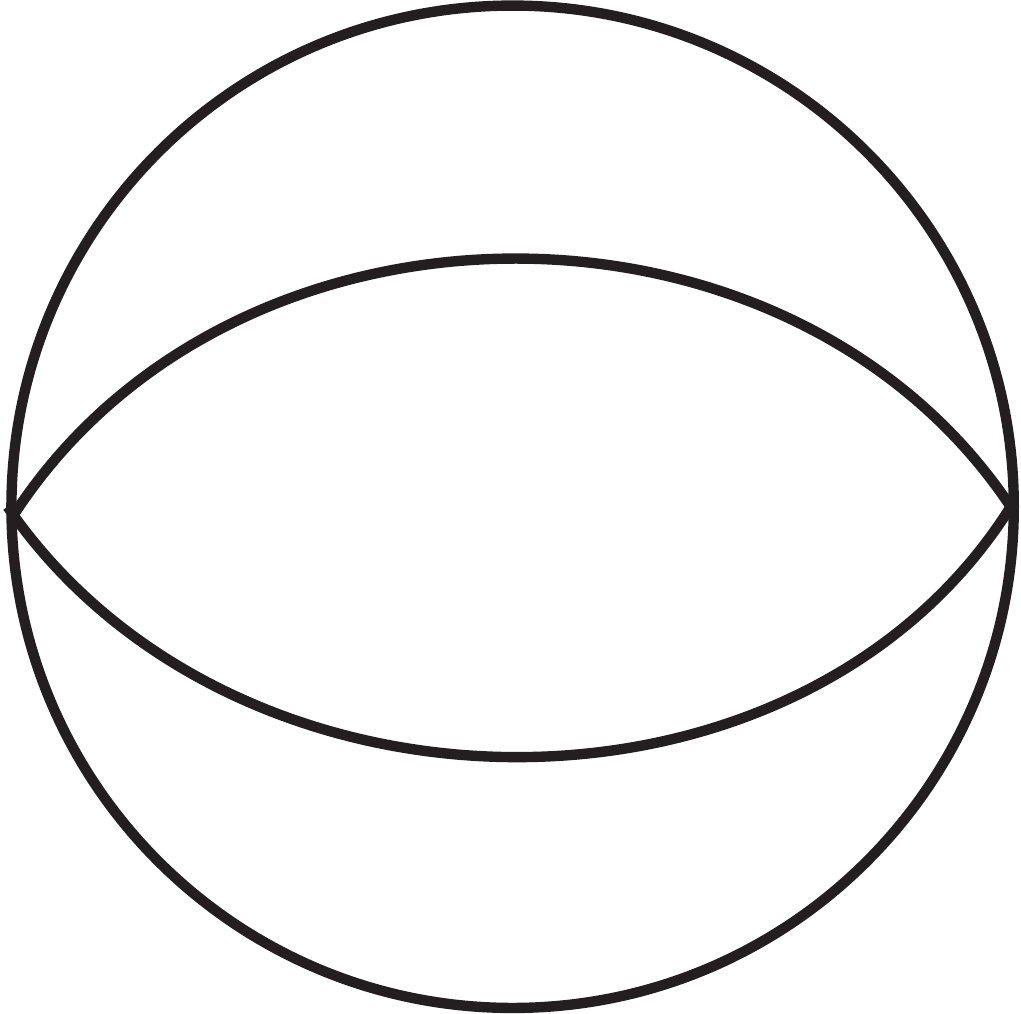}} \qquad
\subfloat[]{\includegraphics[width=0.2\textwidth]{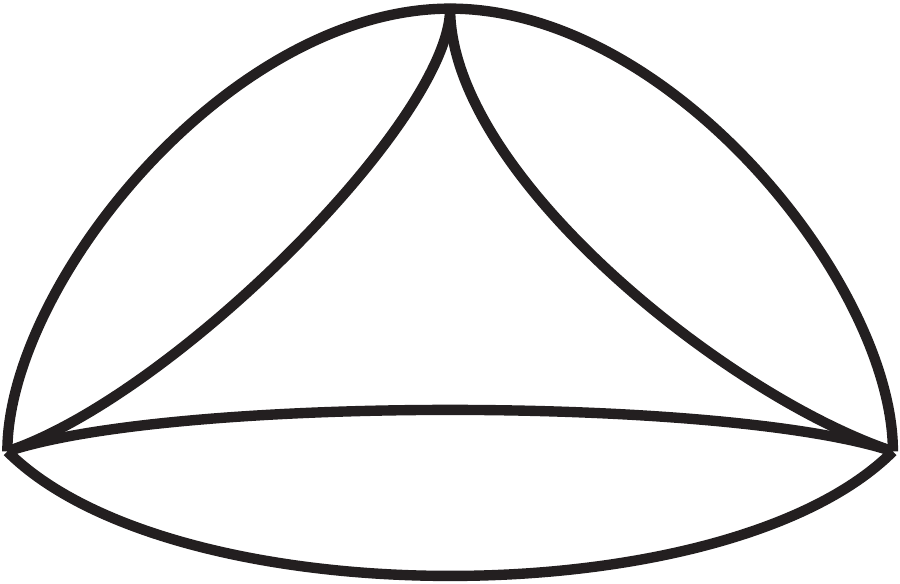}} \qquad
\subfloat[]{\includegraphics[width=0.2\textwidth]{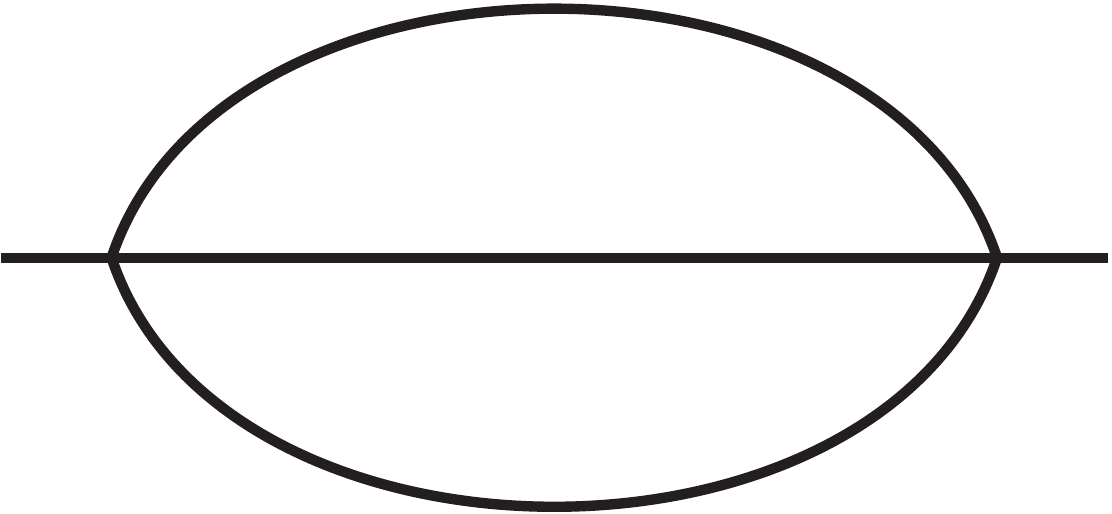}} 
\caption{\label{otherDiagrams} The (a) three-loop and (b) four-loop divergent vacuum diagrams contributing to $\Phi_{LW}$; (c) the two-loop proper self energy obtained from the three-loop vacuum diagram.}
\end{figure}

The proof for UV-finiteness of \eqref{LWnoSSB} is in fact almost trivial: the theory given by \eqref{Hk} and \eqref{Hi} is super-renormalizable in $d=3$, and the usual power-counting argument \cite{Fradkin2021-zu} shows that there is a grand total of three UV-divergent vacuum Feynman diagrams contributing to $\Phi_{LW}$.  The two-loop Hartree-Fock diagram Fig \ref{vacuum2} is treated in the main text, and the other diagrams are shown here in Fig \ref{otherDiagrams}.  The three-loop diagram yields a logarithmic-divergent proper self energy insertion affecting $\alpha$ only, and a corresponding two-loop contribution to $\delta\alpha$ can be computed in perturbation theory.  The four-loop diagram only contains a diverging additive constant that does not affect any physics.  One only needs to explicitly check the UV-finiteness for the two- and three-loop cases.  We thus consider the matter settled, and move on to show that \eqref{LWnoSSB} does yield the desired thermodynamic potential.

The perturbative construction of the original LW formalism can be found in many textbooks, and we recommend the very accessible review given by Eder \cite{Eder2019-il}.  Essentially, one first shows that the non-interacting part of \eqref{LWnoSSB} yields the desired free energy for the non-interacting theory.  And then the interaction is adiabatically turned on, and one proceeds to show that the variation of the LW functional $\Phi_{LW}$ matches the change in free energy.  Note that the whole derivation is in a sense formulated in the ``bare'' theory.  In condensed matter context, the formalism is usually employed to treat an electronic model that either is explicitly defined on a lattice or has a k-space Brillouin zone, and UV-divergence and superfluid order (the two conceptual difficulties here for boson gas) are not even present.  There is no need for renormalization, and no distinction between bare and renormalized theories for the fermionic case.

At the first glance, it seems that if one is to recognize $(\mG_0^{-1} + \delta \alpha)$ as the ``bare free propagator'', then one arrives at \eqref{LWnoSSB} immediately.  But there is a conceptual issue: $\delta\alpha$ is negative and divergent, and this ``bare free propagator'' represents a theory without stable equilibrium; the non-interacting partition function does not even exist, and consequently it cannot be used as a starting point to construct the interacting free energy.  The key point is that $\delta\alpha$ should only be turned on together with the quartic interaction.  The original proof must be adapted to accommodate $\delta \alpha$ as a two-leg interaction vertex, and the rest is otherwise straightforward.

The extension to include the superfluid order \eqref{LWwithSSB} is also technically straightforward with conceptual fine prints.  This time, one may actually desire a negative renormalized value for $\alpha$, but there is of course no such thing as a non-interacting bosonic theory with a negative energy gap.  A prescription for the ``non-interacting theory'' (as good as any other stable choice) is to start with a non-interacting value of $\alpha = 0^{+}$.  Then $\alpha$ is adiabatically taken to the desired negative value, together with the switching-on of $\delta\alpha$ and the quartic interaction.  The other new addition is the effective three-leg interaction vertex in the presence of the superfluid order.  Adapting the proof to include these new elements is not difficult.

\section{Numerical estimation of GL coefficients}

In the main text, we investigate the effective Hamiltonian \eqref{HkPhi} and \eqref{HiPhi} for small $C_1$ and $C_2$.  We will estimate these ratios for $M_x$Bi$_2$Se$_3$.

Estimations for $C_1$ is readily available in the existing literature.  Ref \cite{Zyuzin2017-kn} gave the values of $1/3$ and $1/2$ for two different models.  The present authors estimated from reported $H_{c2}$ anisotropy (for M $=$ Cu) that $(K_2+K_3)/2K_1 \approx 0.6$ \cite{How2020-qj}, and this translates to $C_1 \approx 0.4$.

The ratio $C_2$ involves $K'$ and $K_z$, both much less documented.  $K'$ embodies the breaking of hexagonal symmetry in the basal plane down to trigonal, which is unlikely to be big judging from the crystal structure.  Within the weak-coupling approximation, such a term originates from the anisotropy of the Fermi surface; the normal state Fermi surface was experimentally reconstructed in \cite{Lahoud2013-rl, Almoalem2021-zf} at different doping levels, and it appears to us that neither groups reported a significant departure from full isotropy in the basal plane.  While a numerical estimation is unavailable, we believe the ratio $K'/K_1$ is likely to be much smaller than order unity.

Ref \cite{Fischer2016-gt} adopted $K_z/K_1 = 10^{-2}$ in their model intended to approximate Sr$_2$RuO$_4$.  But this choice is only vaguely guided by the observation that there is ``extreme anisotropy'' in the real material, and of course isn't directly related to $M_x$Bi$_2$Se$_3$.  The ratio of anisotropic $H_{c2}$ (field direction parallel v.s. perpendicular to the basal plane) equals the ratio $\sqrt{K_z/(2K_1 + K_2 + K_3)}$ and $\sqrt{K_z/K_1}$, depending on the relative nematic orientation: see \cite{Hess1989-wi}.  Based on the available $H_{c2}$ measurements \cite{Kriener2011-ha,Pan2016-ni,Smylie2018-ab}, we put $K_z/K_1$ in the neighborhood of $1/10$.

Combining the above observations, we believe $C_2$ is at most of order unity, possibly much less.  This put the effective spin-orbit coefficient $J \lesssim 0.2$, justifying our perturbative treatment.  Conversely, to get $J \approx 1$, assuming the extreme $C_1 = 1$ and keeping the estimation $K_z/K_1 \approx 1/10$, a ratio of $K'/K_1 > 2$ is required, indicating a fairly large anisotropy not seen in the material.

We also note that in the 2D limit of $K_z/K_1 \rightarrow 0$, $C_1 \rightarrow 1$ according to ref \cite{Zyuzin2017-kn}, while $C_2$ diverges.  Therefore the quasi-2D limit is another way to get a large $J$.  This calls for $(K'/K_1)^2 > (K_z/K_1)$ if we seek $J \approx 1$.  Very extreme anisotropy in the $z$-direction is needed if the basal plane is only weakly three-fold anisotropic.  Again this is irrelevant to real materials.

Finally, on a somewhat different track, we want to remark that the chemical potential for superconducting $M_x$Bi$_2$Se$_3$ is reported to be of the order $100$meV or more in \cite{Wray2010-ct}, while its $T_c$ is below $4$K. The ratio $T_c/E_F$ is therefore a small number, and the application of our result (based on GL analysis) to $M_x$Bi$_2$Se$_3$ is well-justified.

\end{document}